\documentstyle[12pt,amsmath]{article}
\numberwithin{equation}{section}

\newcommand{\Tr}{{\rm{Tr}}}

\renewcommand{\theequation}{\arabic{section}.\arabic{equation}}

\newcommand{\grgl}{\:\hbox to -0.2pt{\lower2.5pt\hbox{$\sim$}\hss}
           {\raise3pt\hbox{$>$}}\:}
\newcommand{\klgl}{\:\hbox to -0.2pt{\lower2.5pt\hbox{$\sim$}\hss}
           {\raise3pt\hbox{$<$}}\:}
\def\KK{{\rm I\kern -.2em  K}}
\def\NN{{\rm I\kern -.16em N}}
\def\RR{{\rm I\kern -.2em  R}}
\def\ZZ{Z \kern -.43em Z}
\def\QQ{{\rm \kern .25em
             \vrule height1.4ex depth-.12ex width.06em\kern-.31em Q}}
\def\CC{{\rm \kern .25em
             \vrule height1.4ex depth-.12ex width.06em\kern-.31em C}}
\def\ZZZ{Z\kern -0.31em Z}
\def\Lc{{\cal L}}

\begin{document}
\begin{titlepage}

\quad\\
\vspace{1.8cm}
\begin{center}
{\bf\LARGE Quantum correlations}\\
\bigskip
{\bf\LARGE from incomplete classical statistics}\\
\vspace{1cm}
C. Wetterich\footnote{e-mail: C.Wetterich@thphys.uni-heidelberg.de}\\
\bigskip
Institut  f\"ur Theoretische Physik\\
Universit\"at Heidelberg\\
Philosophenweg 16, D-69120 Heidelberg\\
\vspace{3cm}
\today
%{\bf Abstract:}\\
%\parbox[t]{\textwidth}{}
\begin{abstract}
We formulate incomplete classical statistics for situations
where the knowledge about the probability distribution
outside a local region is limited.
The information needed to compute expectation values of local observables can
be collected in a quantum mechanical state vector, whereas further statistical
information about the probability distribution outside the local region
becomes irrelevant. The translation of the available information between neighboring
local regions is expressed by a Hamilton operator. A quantum mechanical operator
can be associated to each local observable, such that expectation values of ``classical''
observables can be computed by the usual quantum mechanical rules.
The requirement that correlation
functions should respect equivalence relations for local obeservables induces a
non-commutative product in classical statistics, in complete correspondence to
the quantum mechanical operator product. We also discuss the issue of interference
and the complex structure of quantum mechanics within our classical statistical setting.
\end{abstract}
\end{center}\end{titlepage}
\newpage

\section{Incomplete classical statistics}

Following Einstein, Rosen and Podolski \cite{EPR}, many physicists
have asked if it could be possible to derive quantum mechanics from
an underlying classical statistical system. The motivation is to find
an explanation of the ``why'' of the basic principles of quantum
mechanics like the formulation in terms of
states and non-commuting operators, the superposition of
``probability amplitudes'' and the associated interference effects.
Such a derivation from an underlying theory could
also open the way to possible generalizations of quantum mechanics
and to the formulation of further tests of its basic
principles \cite{SW}, \cite{GT}. For a large class of classical statistical
systems with certain locality properties it has been
shown \cite{Bell} that classical and quantum statistics lead to
a different behavior of measurable correlation functions.
The experimental verification of the predictions of quantum
mechanics is unequivocal. Any attempt to derive quantum mechanics
from classical statistics has therefore to circumvent the assumptions
of Bell's inequalities for the classical system and should reproduce
the predictions from quantum mechanical interference. Our formulation
of incomplete classical statistics with infinitely many degrees of
freedom is of this type.

The understanding of classical statistical
systems with infinitely many degrees of freedom has made tremendous
progress in the past decades. Important links to quantum mechanics
have been established by the use of path integrals \cite{Feynman} for the
description of the quantum mechanical evolution.
This process has diminished considerably the distance between
classical statistics and quantum mechanics or quantum
field theory \cite{1a}. One wonders if
quantum mechanics cannot be understood as a particular structure of
classical statistical systems with infinitely many degrees of
freedom or, in other words, if it can be derived from general
statistics \cite{GenStat}. Within general statistics  the notions of
distance, geometry and topology have already been formulated in terms of
properties of   correlation functions \cite{Geom}.  The question arises if
the notions of time and   quantum mechanical evolution can find their
origin within the same framework.

The formulation of the basic partition function for classical
statistical   systems with  infinitely many degrees of freedom uses
implicitly an assumption   of ``completeness of the statistical
information''. This means that we assign   a probability to everyone
of the infinitely many configurations. The   specification of the
probability distribution contains therefore an ``infinite   amount of
information''. This contrasts with the simple observation that only a
finite amount of information is available in practice for the
computation of   the outcome of any physical measurement.
A concentration on measurable quantities suggests that the assumption
of completeness of the statistical information may have to be abandoned.
In this note
we explore consequences of ``incomplete statistics'' which deals
with situations where only partial information
about the probability distribution is available. In   particular, we
consider extended systems for which only local information about   the
probability distribution is given. We will see that the quantum
mechanical   concepts of states, operators, evolution and interference
emerge naturally in this setting.

The outcome of the present work is still far from reaching a stage where
quantum mechanics can be derived from a classical statistical setting.
The main obstacle remains the characteristic complex structure of quantum
mechanics which is not yet implemented in a satisfactory way. Nevertheless,
we find it interesting to observe how other structures which were believed to be
characteristic for quantum mechanics, like a non-commutative product or the effect
of interference, are actually already present in classical statistics. The
formulation of incomplete statistics will be very useful in order to motivate
how the quantum mechanical structures arise naturally in classical statistics.
The notion of incompleteness is, however, not crucial for the existence of these
structures. Quantum mechanical structures are present in ``standard'' (complete)
classical statistics as well.

As an example, let us consider a classical statistical system where the
infinitely   many degrees of freedom  $\varphi_n\ (n\in\ZZ)$ are
ordered  in an infinite   chain. We concentrate on a ``local region''
$|\tilde n|<\bar n$ and  assume   that the probability distribution
$p[\varphi]$ has a ``locality property''
in the sense that the relative probabilities
for any two configurations of the ``local
variables'' $\varphi_{\tilde n}$ are independent of the values that
take the  ``external variables''
$\varphi_m$ with $|m|>\bar n$. Furthermore, we
assume that the probability distribution for the $\varphi_{\tilde n}$
is known for given values of the variables $\varphi_{\bar n},
\varphi_{-\bar n}$ at the border of the local interval.
As an example, we may consider a probability distribution
\begin{eqnarray}\label{1.1}
p[\varphi]&=&p_>[\varphi_{m\geq\bar n}]p_0[\varphi_{-\bar n\leq\tilde n
\leq\bar n}]p_<[\varphi
_{m\leq-\bar n}]\\
p_0[\varphi]&=&\exp\left\{-\sum_{|\tilde n|<\bar n}\left[\frac{\epsilon}{2}\mu^2
\varphi^2_{\tilde n}+\frac{\epsilon}{8}\lambda\varphi^4_{\tilde n}
+\frac{M}{2\epsilon}(\varphi_{\tilde n}-\varphi_{\tilde n-1})^2\right]
-\frac{M}{2\epsilon}(\varphi_{\bar n}-\varphi_{\bar n-1})^2\right\}
\nonumber\end{eqnarray}
where $p_>$ and $ p_<$ are only constrained by the overall normalization
of $p[\varphi]$ and we will consider the limit $\epsilon\to 0$.
This
statistical system cannot be reduced to a system with a finite number
of degrees of freedom since the probability for the occurrence of
specific values of the ``border variables'' $\varphi_{\bar
n},\varphi_{-\bar n}$ depends on the values of the external
variables  $\varphi_m$ and their probability distribution. The
statistical information about this system is incomplete since we will
not specify the probability distribution
$p_>p_<$ for the external variables
$\varphi_m$ completely.

Local observables are constructed from the local variables
$\varphi_{\tilde n}$. As usual, their expectation values are computed
by ``functional integrals'' where the probability distribution
$p[\varphi]$ appears as a weight factor.
Incomplete statistics deal with the expectation values of local observables.
Thereby the questions that may be asked reflect the incompleteness
of the statistical information. As a first question, we ask what are
the conditions on $p_>p_<$ which lead to $\tilde n$-independent
expectation values $<\varphi^p(\tilde n)>$ for all or some integer
values of $p$? Furthermore, are there any restrictions on the allowed
values of $\tilde n$-independent expectation values $<\varphi^p(\tilde n)>$?
The answer to these questions, given in sect. 4, may surprise: indeed,
probability distributions $p_>,p_<$ that are consistent with
$\tilde n$-independent  $<\varphi^p(\tilde n)>$ are possible only
for certain discrete values for $<\varphi^p(\tilde n)>$. These values
correspond precisely to the expectation values of the operators
$\hat Q^p$ in the stationary states of quantum mechanics. Here our example
(\ref{1.1}) corresponds to a one-dimensional particle with mass
$M$ in an unharmonic potential $V(\hat Q)=\frac{\mu^2}{2}\hat Q^2+\frac
{\lambda}{8}\hat Q^4$. The appearance of the quantum-mechanical
discreteness in a classical statistical system is striking and we will
attempt to understand its origin and the deep connection to quantum
mechanics in more detail in this note.

We next ask more generally what is the
minimal amount of information about the probability distribution for
the external variables $\varphi_m$
which is necessary for a computation of
expectation values of local observables. One finds that this
information can be summarized in ``states'' $|\psi\},\{\psi|$
that can be represented as ordinary functions $\{\psi(\varphi_{\bar n})|$,
$|\psi(\varphi_{-\bar n})\}$. Since these functions depend each only
on one variable, the
specification of the states contains much less information than
the full probability  distribution $p_>p_<$ which depends on
infinitely many variables $\varphi_{m\geq\bar n}$, $\varphi_{m\leq-\bar n}$.
The states
contain the minimal information for ``local questions'' and  are
therefore the
appropriate quantities for our formulation of incomplete
statistics.
We will see in sect. 7 that any further information about the probability
distributions $p_>[\varphi_m],p_<[\varphi_m]$ beyond the one contained in the
state vectors is actually irrelevant for the computation of expectation values
of local observables.

The expectation values of all local observables can be
computed from the know\-ledge of the local probability distribution and
the states $|\psi\}$ and $\{\psi|$. For this computation one
associates to every local observable $A[\varphi]$ an appropriate operator
$\hat A$ and finds the prescription familiar from quantum mechanics
\begin{equation}\label{1.2}
\langle A[\varphi]\rangle =\{\psi|\hat A|\psi\}\end{equation}
There is a unique mapping $A[\varphi]\to\hat A$ for every
local observable which can be expressed in terms
of an appropriate functional integral. We find that for simple observables $A[\varphi]$ the
operators $\hat A$ correspond precisely to familiar operators in quantum
mechanics. For example, the observable $\varphi(\tilde n)$
can be associated to the operator $\hat Q(\tau)$ in the Heisenberg
picture where time is analytically continued, $\tau=it$, and $\tilde n
=\tau/\epsilon$.

Local correlation functions involving derivatives may be ambiguous in
the continuum limit. This problem is well known in functional
integral formulations of quantum field theories. We show how to
avoid this problem by defining correlations in terms of equivalence
classes of observables. In fact, two
observables $A_1[\varphi],A_2[\varphi]$ can sometimes be represented
by the same operator $\hat A$. In this case $A_1[\varphi]$
and $A_2[\varphi]$ are equivalent since they cannot be
distinguished by their expectation values for arbitrary states.
They have the same expectation values for all possible probability distributions.
We define a product between equivalence classes of
observables which can be associated to the  product of
operators. For example, we associate to a
correlation $\varphi(\tilde n_1)\circ \varphi(
\tilde n_2)$ the product $\hat Q(\tau_1)\hat Q(\tau
_2)$ which is not commutative. It is again striking how the non-commutativity
of quantum mechanics arises directly from the question what
are meaningful correlation functions in incomplete statistics!
Indeed,
equivalent observables
should lead to equivalent correlations.
We will see (sect. 6) that this points to
a non-commutative definition of the correlation
$\varphi(\tilde n_1)\circ\varphi(\tilde n_2)$. In sect. 7 we argue that
the ``quantum correlation'' based on $\varphi_1\circ\varphi_2$
can actually be formulated quite generally in classical statistics
(not necessarily incomplete) and has better ``robustness
properties'' as compared to the usual classical correlation.

We also may ask what can be learned for expectation values like
$<\varphi^p(\tilde n)>$ if the states are specified by a sufficient
number of expectation values at some given location $\tilde n_0$
(e.g. $<\varphi^p(\tilde n_0)>$, $<(\varphi(\tilde n_0+1)-\varphi(\tilde
n_0))^{p'}>$ etc.). We find that operators and states can be transported
from one site to the next by an
evolution operator
$\hat U$ in complete analogy to quantum mechanics. This
introduces once more a crucial concept of quantum
mechanics in the formulation of incomplete statistics, namely the Hamilton
operator. As one may expect, it corresponds to the transfer matrix in classical
statistics.

Finally, we ask in sect. 8
if the characteristic interference of quantum mechanics
also finds its equivalent in incomplete statistics.
We explicitly construct classical probability distributions
that correspond to linear superpositions of states
$\alpha_1|\psi_1\}+\alpha_2|\psi_2\}$ and show the corresponding
interference behavior. We also describe probability distributions
leading to a formulation in terms of density matrices that do not
correspond to pure states.
We conclude in sect. 9 that many fundamental features of quantum
mechanics are shown to arise directly from the formulation of
incomplete statistics. Inverting our procedure we find quantum
mechanical states which admit a functional integral description. This goes
beyond the usual description of the vacuum or thermal equilibrium state.

There remains one crucial feature of quantum mechanics, namely its complex
structure or the issue of ``phases'', that has not yet emerged in a
satisfactory way from classical statistics. It is closely related to the
difference between ``real time'' in quantum mechanics and ``Euclidean time''
in classical statistics or
the issue of analytical continuation. We describe our (incomplete) attempts in this
direction in two appendices. The invariance of the probability distribution under
a reflection $\varphi_n\to\varphi_{-n}$ can be associated with a complex
structure. With respect to this complex structure all quantities that are odd under
reflections become purely imaginary. Then the evolution operators $\hat U$ are
unitary and $\{\psi|$ is complex conjugate to $|\psi\}$. Also $\tau=\epsilon n$ is
imaginary. It is not clear, however, how the state vectors as functions of $\varphi_n$
(at fixed $n$) should be analytically continued without loosing the probability
interpretation of the functional integral.

\bigskip
\section{States and operators}
\setcounter{equation}{0}

\medskip
Consider a discrete ordered set of continuous variables
$\varphi_n\equiv\varphi(\tau)$, $\tau=\epsilon n$, $n\in\ZZ$ and a
normalized probability distribution $p(\{\varphi_n\})\equiv
p[\varphi]=\exp(-S[\varphi])$ with $\int D\varphi
e^{-S[\varphi]}\equiv \prod_n(\int
^\infty_{-\infty}d\varphi_n)p[\varphi]=1$.  We will assume that the
action $S$ is local in a range $-\bar\tau<\tau<\bar\tau$, i.e.
\begin{eqnarray}\label{2.1}
S&=&-\ln p=\int^{\bar\tau}_{-\bar\tau}d\tau'{\Lc}(\tau')
+S_>(\bar\tau)+S_<(-\bar\tau)\nonumber\\
{\Lc}(\tau')&=&V(\varphi(\tau'),\tau')+
\frac{1}{2}Z(\tau')(\partial_{\tau'}\varphi(\tau'))^2\end{eqnarray}
Here we have used a continuum notation $(n_{1,2}=\tau_{1,2}/\epsilon)$
which can be translated into a discrete language by
\begin{eqnarray}\label{2.2}
&&\int^{\tau_2}_{\tau_1}d\tau'{\Lc}(\tau')=\epsilon\sum^{n_2-1}
_{n=n_1+1}{\Lc}_n+\frac{\epsilon}{2}\left[
V_{n_2}(\varphi_{n_2})+V_{n_1}(\varphi_{n_1})\right]\nonumber\\
&&+\frac{\epsilon}{4}\left[Z_{n_2}\left(\frac{\varphi_{n_2}
-\varphi_{n_2-1}}
{\epsilon}\right)^2+Z_{n_1}\left(\frac{\varphi_{n_1+1}
-\varphi_{n_1}}{\epsilon}\right)^2\right] \end{eqnarray}
with
\begin{equation}\label{2.2a}
{\cal L}_n=V_n(\varphi_n)+\frac{Z_n}{4\epsilon^2}\{(\varphi_{n+1}
-\varphi_n)^2+(\varphi_n-\varphi_{n-1})^2\}\end{equation}
This corresponds to a discrete derivative
\begin{eqnarray}\label{2.3}
(\partial_\tau\varphi(\tau))^2&=&\frac{1}{2}
\left\{\left(\frac{\varphi
(\tau+\epsilon)-\varphi(\tau)}{\epsilon}\right)^2+ \left(\frac{\varphi
(\tau)-\varphi(\tau-\epsilon)}{\epsilon}\right)^2\right\}\nonumber\\
&=&\frac{1}{2\epsilon^2}\left\{\left(\varphi_{n+1}-\varphi_n\right)^2
+\left(\varphi_n-\varphi_{n-1}\right)^2\right\}.\end{eqnarray}
The boundary
terms in eq. (\ref{2.2}) are chosen such that $S_>(\bar\tau)$ is
independent of all $\varphi(\tau')$ with $\tau'<\bar\tau$ whereas
$S_<(-\bar\tau)$ only depends on $\varphi(\tau'\leq-\bar\tau)$. Except
for the overall normalization of $p$ no additional assumptions about
the form of $S_>(\bar\tau)$ and $S_<(-\bar\tau)$ will be made. In case
of $S$ being local also at $\bar\tau$ we note that $S_>(\bar\tau)$
contains a term $\frac{\epsilon}{2}[V(\varphi(\bar\tau),\bar\tau)+
V(\varphi(\bar\tau+\epsilon),\bar\tau+\epsilon)]+\frac{\epsilon}
{4}(Z(\bar\tau)+Z(\bar\tau+\epsilon))\left(\frac{
\varphi(\bar\tau+\epsilon)-\varphi(\bar\tau)}{\epsilon}\right)^2$,
which involves a product $\varphi(\bar\tau+\epsilon)\varphi(\bar\tau)$
and therefore links the variables with $\tau>\bar\tau$ to the ones
with $\tau\leq \bar\tau$.

We are interested in local observables $A[\varphi;\tau]$ which depend
only on those $\varphi(\tau')$ where $\tau-\frac{\delta}
{2}\leq\tau'\leq\tau+\frac{\delta}{2}$.  (We assume
$-\bar\tau<\tau-\frac{\delta}{2}$, $\bar\tau> \tau+\frac{\delta}{2}$.)
As usual, the expectation value of $A$ is
\begin{equation}\label{2.4}
<A(\tau)>=\int
D\varphi A[\varphi;\tau]e^{-S[\varphi]} \end{equation}
As mentioned in the introduction, our
investigation concerns the question what we can learn about
expectation values of local observables and suitable   products
thereof in a situation where we have no or only partial information
about $S_>(\bar\tau)$ and $S_<(-\bar\tau)$.  It seems obvious that the
full information contained in $S$ is not needed if only expectation
values of local observables of the type (\ref{2.4}) are to be computed.
On the
other hand, $<A(\tau)>$ cannot be completely independent of $S_>
(\bar\tau)$ and $S_<(-\bar\tau)$ since the next neighbor interactions
(\ref{2.2}) relate ``local variables'' $\varphi(\tau-\frac{\delta}{2}<\tau'
<\tau+\frac{\delta}{2})$ to the ``exterior variables'' $\varphi(\tau'>\bar\tau)$ and
$\varphi(\tau'<-\bar\tau)$.

In order to establish the necessary amount of information needed from
$S_>(\bar\tau)$ and $S_<(-\bar\tau)$ we first extend $S_>$ and $S_<$
to values $|\tau|<\bar\tau$
\begin{equation}\label{2.5a}
S_<(\tau_1)=S_<(-\bar\tau)+\int^{\tau_1}
_{-\bar\tau}d\tau'{\Lc}(\tau')\ ,\
 S_>(\tau_2)=S_>(\bar\tau)+\int^{\bar\tau}_{\tau_2}
d\tau'{\Lc}(\tau')\end{equation}
where we note the general identity
\begin{equation}\label{2.5}
S_>(\tau)+S_<(\tau)=S \end{equation}
The expectation value (\ref{2.4}) can be written as
\begin{eqnarray}\label{2.6} &&<A(\tau)>=\int
d\varphi(\tau+\frac{\delta}{2})\int
d\varphi(\tau-\frac{\delta}{2})\int
D\varphi_{(\tau'>\tau+\frac{\delta}
{2})}e^{-S_>(\tau+\frac{\delta}{2})}\\
&&\int
D\varphi_{(\tau-\frac{\delta}{2}<\tau'<\tau+\frac{\delta}{2})}
A[\varphi;\tau]\exp\{-\int^{\tau+\frac{\delta}{2}}
_{\tau-\frac{\delta}{2}}d\tau''{\Lc}(\tau'')\}\int
D\varphi_{(\tau'<\tau-\frac{\delta}{2})}e^{-S_<(\tau-
\frac{\delta}{2})}\nonumber\end{eqnarray}
This suggests the introduction of the
``states''
\begin{eqnarray}\label{2.7}
&&|\psi(\varphi(\tau-\frac{\delta}{2});\
\tau-\frac{\delta}{2})\} =\int
D\varphi_{(\tau'<\tau-\frac{\delta}{2})}e^{-S_<(\tau
-\frac{\delta}{2})}\nonumber\\
&&\{\psi(\varphi(\tau+\frac{\delta}{2});\ \tau+\frac{\delta}{2})|
=\int D\varphi_{(\tau'>\tau+\frac{\delta}{2})}
e^{-S_>(\tau+\frac{\delta}{2})}\end{eqnarray}
and the operator
\begin{eqnarray}\label{2.8}
&&\hat A_\delta(\varphi(\tau+\frac{\delta}{2}),\ \varphi(\tau-
\frac{\delta}{2});\ \tau)\nonumber\\ &=&\int
D\varphi_{(\tau-\frac{\delta}{2}<\tau'<\tau+\frac{\delta}{2})}
A[\varphi;\ \tau]\exp\left\{-\int^{\tau+\frac{\delta}{2}}
_{\tau-\frac{\delta}{2}}d\tau''{\Lc}(\tau'')\right\}\end{eqnarray}
We note that
$|\psi\}$ is a function of $\varphi(\tau- \frac{\delta}{2})$ since the
latter appears in $S_<(\tau-\frac{\delta} {2})$ and is not included in
the (``functional'') integration (\ref{2.7}). Similarly, $\{\psi|$
depends on $\varphi(\tau +\frac{\delta}{2})$ whereas $\hat A$ is a
function of the two variables $\varphi(\tau +\frac{\delta}{2})$ and
$\varphi(\tau-\frac{\delta}{2})$. Using a notation where $|\psi\}$ and
$\{\psi|$ are interpreted as (infinite dimensional) vectors and $\hat
A$ as a matrix, one has
\begin{eqnarray}\label{2.9}
&&<A(\tau)>=\{\psi(\tau+\frac{\delta}{2})\hat A_\delta
(\tau)\psi(\tau-\frac{\delta}{2})\}\nonumber\\ &&\equiv\int
d\varphi_2\int d\varphi_1\{\psi(\varphi_2;\tau+\frac{\delta} {2})|\hat
A_\delta(\varphi_2,\varphi_1;\tau)|
\psi(\varphi_1;\tau-\frac{\delta}{2})\}\end{eqnarray}
This form resembles
already the well-known prescription for expectation values of
operators in quantum mechanics.  In contrast to quantum mechanics
eq. (\ref{2.9}) still involves, however, two different state vectors.

The mapping $A[\varphi;\ \tau]\to\hat A_\delta(\tau)$ can be computed
(cf. eq. (\ref{2.8})) if ${\Lc}(\tau')$ is known for
$|\tau'|<\bar\tau$. The only information needed from $S_>(\bar\tau)$
and $S_<(-\bar\tau)$ is therefore contained in the two functions
$\{\psi(\varphi)|$ and $|\psi(\varphi)\}$! The specification of these
states (wave functions) at $\bar\tau$ and $-\bar\tau$ and of ${\Lc}
(|\tau|<\bar\tau)$ completely determines the expectation values of
{\em all} local observables!

We will see below the close
connection to the states in quantum mechanics. In our context we
emphasize that for any given $S$ these states can be computed as well
defined functional integrals (\ref{2.7}). Due to eq. (\ref{2.5}) they obey
the normalization
\begin{equation}\label{2.10} \{\psi(\tau)\psi(\tau)\} \equiv\int
d\varphi\{\psi(\varphi;\tau)||\psi(\varphi;\tau)\}=1\end{equation}
Incomplete statistics explores statements that can be made
for local observables and appropriate products thereof  without using information about $S_>$ or $S_<$ beyond the
one contained in the states $|\psi\}$ and $\{\psi|$.

\bigskip
\section{ Evolution in Euclidean time}
\setcounter{equation}{0}
\medskip
For a ``locality interval'' $\delta>0$ the expression (\ref{2.9})) involves
states at different locations or
``Euclidean times'' $\tau+\frac{\delta} {2}$ and
$\tau-\frac{\delta}{2}$. We aim for a formulation where only states at
the same $\tau$ appear. We therefore need the explicit mapping from
$|\psi(\tau-\frac{\delta}{2})\}$ to a reference point $|\psi(\tau)\}$
and similar for $\{\psi(\tau+\frac{\delta}{2})|$. This mapping
should also map $\hat A_\delta$ to a suitable operator such that the
structure (\ref{2.9}) remains preserved.
The dependence of states and operators on the
Euclidean time $\tau$ is described by evolution operators
$(\tau_2>\tau_1,\ \tau_2>\tau_f,\
\tau_i>\tau_1,\tau_f=\tau+\frac{\delta}{2},\
\tau_i=\tau-\frac{\delta}{2}))$
\begin{eqnarray}\label{3.1}
&&|\psi(\tau_2)\}=\hat
U(\tau_2,\tau_1)|\psi(\tau_1)\}\nonumber\\
&&\{\psi(\tau_1)|=\{\psi(\tau_2)|\hat U(\tau_2,\tau_1)\nonumber\\
&&\hat A(\tau_2,\tau_1)=\hat U(\tau_2,\tau_f)\hat A
(\tau_f,\tau_i)\hat U(\tau_i,\tau_1) \end{eqnarray}
or differential operator
equations $(\epsilon\to0)$;
\begin{eqnarray}\label{3.2}
&&\partial_\tau|\psi(\tau)\}=-\hat H(\tau)|\psi(\tau)\}\nonumber\\
&&\partial_\tau\{\psi(\tau)|=\{\psi(\tau)|\hat H(\tau)\nonumber\\
&&\partial_{\tau_i}\hat A_\delta(\tau_f,\tau_i)= \hat
A_\delta(\tau_f,\tau_i)\hat H(\tau_i)\nonumber\\
&&\partial_{\tau_f}\hat A_\delta(\tau_f,\tau_i)=-\hat H(\tau_f) \hat
A_\delta(\tau_f,\tau_i).\end{eqnarray}
The evolution operator has an explicit
representation as a functional integral
\begin{equation}\label{3.3} \hat
U(\varphi(\tau_2),\varphi(\tau_1);\ \tau_2,\tau_1) =\int
D\varphi_{(\tau_1<\tau'<\tau_2)}\exp\left\{
-\int^{\tau_2}_{\tau_1}d\tau''{\Lc}(\tau'')\right\}\end{equation}
and obeys the
composition property $(\tau_3>\tau_2>\tau_1)$
\begin{equation}\label{3.4} \hat
U(\tau_3,\tau_2)\hat U(\tau_2,\tau_1)=\hat U(\tau_3,\tau_1)\end{equation}
with
\begin{equation}\label{3.5} \hat U(\varphi_2,\varphi_1;
\tau,\tau)=\delta(\varphi_2-\varphi_1)\end{equation}
It can therefore be composed
as a product of transfer matrices or ``infinitesimal'' evolution
operators
\begin{equation}\label{3.6}
\hat U(\tau+\epsilon,\tau)=e^{-\epsilon\hat
H(\tau+\frac{\epsilon}{2})}\end{equation}

In case of translation symmetry for the local part of the probability
distribution, i. e. for $V$ and $Z$ independent of $\tau$,  we note the
symmetry in $\varphi_1\leftrightarrow\varphi_2$
\begin{equation}\label{3.7} \hat
U(\tau+\epsilon,\tau)=\hat U^T(\tau+\epsilon,\tau)\ ,\ \hat H(\tau
+\frac{\epsilon}{2})= \hat H^T(\tau+\frac{\epsilon}{2})=\hat H\end{equation}
In this case the real symmetric matrix $\hat H$ has real eigenvalues
$E_n$. Then the general solution of the differential equation
(\ref{3.2}) may be written in the form
\begin{equation}\label{3.8}
|\psi(\tau)\}=\sum_n\psi_0^{(n)}e^{-E_n\tau},\
\{\psi(\tau)|=\sum_n\bar\psi_0^{(n)}e^{E_n\tau}\end{equation}
where $\psi_0^{(n)}$ and $\bar\psi_0^{(n)}$ are eigenvectors
of $\hat H$ with eigenvalues $E_n$. Here we recall that the construction
(\ref{2.7}) implies that $|\psi\}$ and $\{\psi|$ are real positive functions
of $\varphi$ for every $\tau$. This restricts the allowed values of the
coefficients $\psi^{(n)}_0,\bar{\psi}^{(n)}_0$.

We next want to compute the explicit form of the Hamilton
operator $\hat H$. It is fixed uniquely by the functional
integral representation (\ref{3.3}) for $\hat U$. In order to
obey the defining equation (\ref{3.6}), the  Hamilton
operator $\hat H$ must fulfill for arbitrary $|\psi(\varphi)\}$ the
relation (with
$Z=Z(\tau+\frac{\epsilon}{2})=\frac{1}{2}(Z(\tau+\epsilon) +Z(\tau))$
\begin{eqnarray}\label{3.9}
&&\int d\varphi_1\hat
H(\varphi_2,\varphi_1)|\psi(\varphi_1)\}
=-\lim_{\epsilon\to0}\frac{1}{\epsilon}\Big\{ \int d\varphi_1\\
&&\qquad\exp\left[-\frac{\epsilon}{2}(V(\varphi_2)
+V(\varphi_1))-\frac{Z}{2\epsilon}(\varphi_2-\varphi_1)^2\right]
|\psi(\varphi_1)\}-|\psi(\varphi_2)\}\Big\}\nonumber \end{eqnarray} The solution
of this equation can be expressed in terms of the operators
\begin{eqnarray}\label{3.10}
&&\hat Q(\varphi_2,\varphi_1)=\varphi_1\delta(\varphi_2-\varphi_1)
\nonumber\\ &&\hat
P^2(\varphi_2,\varphi_1)=-\delta(\varphi_2-\varphi_1)
\frac{\partial^2}{\partial \varphi_1^2}\end{eqnarray}
as
\begin{equation}\label{3.11}
\hat H(\tau)=V(\hat Q,\tau)+\frac{1}{2Z(\tau)} \hat P^2\end{equation}
This can be
established\footnote{The Hamilton operator can also be defined for
small nonzero $\epsilon$ where one has $e^{-\epsilon\hat
H}=\left(\frac{2\pi\epsilon}{Z}\right)^{1/2}
e^{-\frac{\epsilon}{2}V(\hat Q,\tau+\epsilon)}e^{-\epsilon \hat
P^2/2Z}e^{-\frac{\epsilon}{2}V(\hat Q,\tau)}$. We have absorbed the
prefactor by a constant shift in $V$, i.e.  $V\to
V-\frac{1}{2\epsilon}\ln\frac{2\pi\epsilon}{Z}$. If appropriate, one
may use the symmetric form where $\partial_1^2
\to\frac{1}{4}(\vec\partial_1-{\stackrel{\leftarrow}{\partial_2}})^2,
\partial_i=\partial/\partial\varphi_i.$} by using under the
$\varphi_1$-integral the replacement
\begin{equation}\label{3.12}
e^{-\frac{Z}{2\epsilon}(\varphi_2-\varphi_1)^2}
\to\left(\frac{2\pi\epsilon}{Z}\right)^{1/2}\delta
(\varphi_2-\varphi_1)\exp\left(\frac{\epsilon}{2Z}\frac{\partial^2}
{\partial\varphi^2_1}\right)\end{equation}
which is valid by partial integration
if the integrand decays fast enough for $|\varphi_1|\to\infty$. We
note that the operators $\hat Q$ and $\hat P^2$ do not commute, e.g.
\begin{equation}\label{3.13} [\hat P^2,\hat
Q](\varphi_2,\varphi_1)=-2\delta(\varphi_2-\varphi_1)
\frac{\partial}{\partial\varphi_1}\end{equation}

The Hamilton operator can be used in order to establish the existence
of the inverse of the ``infinitesimal'' evolution operator, $\hat
U^{-1}(\tau+\epsilon,\tau)=e^{\epsilon\hat H
(\tau+\frac{\epsilon}{2})}$. Then the inverse $\hat U^{-1}
(\tau_2,\tau_1)$ is defined by the multiplication of ``infinitesimal''
inverse evolution operators, and we can extend the composition
property (\ref{3.4}) to arbitrary $\tau$ be defining for
$\tau_2<\tau_1$
\begin{equation}\label{3.14} \hat U(\tau_2,\tau_1)=\hat
U^{-1}(\tau_1,\tau_2)\end{equation}
(For a given dependence of $\hat U$ on the
variables $\tau_2$ and $\tau_1$ the matrix $\hat U(\tau_1,\tau_2)$
obtains from $\hat U(\tau_2,\tau_1)$ by a simple exchange of the
arguments $\tau_1$ and $\tau_2$.) Using eq. (\ref{3.1}), this allows us to
write the expectation value of a local observable in a form involving
states at the same $\tau$-variable
\begin{equation}\label{3.15}
<A(\tau)>=\{\psi(\tau)\hat U(\tau,\tau+\frac{\delta}{2}
)\hat A_\delta(\tau)\hat U(\tau-\frac{\delta}{2},\tau)\psi(\tau)\}\end{equation}

\section{Stationary observables}
\setcounter{equation}{0}
 At this point we can already answer the question posed in the introduction:
What are the conditions for the expectation values $\varphi^p(\tau)$ to be
$\tau$-independent for all $p$ or some given $p$? Since the observables
$\varphi(\tau)$ have support only at one point $\tau$, we can choose
$\delta=0$ such that
\begin{equation}\label{44.1}
<\varphi^p(\tau)>=\{\psi(\tau)\hat Q^p\psi(\tau)\}
=\int d\varphi\{\psi(\varphi;\tau)|\varphi^p|\psi(\varphi;\tau)\}\end{equation}
Using the explicit solution of the evolution equation (\ref{3.8}), this yields
for translation symmetric $V$ and $Z$
\begin{equation}\label{44.2}
<\varphi^p(\tau)>=\int d\varphi\sum_{n,n'}\bar a_n\chi_n
(\varphi)e^{E_n\tau}\varphi^pe^{-E_{n'}\tau}a_{n'}\chi_{n'}(\varphi)\end{equation}
where the functions $\chi_n(\varphi)$ are an orthonormal set of basis
vectors. We choose them to be eigenfunctions of the Hamiltonian
\begin{equation}\label{44.3}
\hat H\chi_n(\varphi)=E_n\chi_n(\varphi)\end{equation}
where $\hat H$ is interpreted as a differential operator
which takes for $Z=M$ the familiar form
\begin{equation}\label{44.4}
\hat H=V(\varphi)-\frac{1}{2M}\frac{\partial^2}{\partial\varphi^2}\end{equation}
We specialize to $V(\varphi)=\frac{1}{2}\mu^2\varphi^2+
\frac{1}{8}\lambda\varphi^4,\ \lambda>0$,
which corresponds to an unharmonic oscillator (cf. eq. (\ref{1.1})).
The functions $\chi_n(\varphi)$ are, of course, precisely
the stationary solutions of the Schr\"odinger equation for a
particle in a one-dimensional potential $V(\varphi)$.

Defining
\begin{equation}\label{44.5}
K^p_{nn'}=\int d\varphi \chi_n(\varphi)\varphi^p\chi_{n'}(\varphi)\end{equation}
one obtains
\begin{equation}\label{44.6}
<\varphi^p(\tau)>=\sum_{n,n'}\bar a_na_{n'}K^p_{nn'}e^{(E_n-E_{n'})
\tau}\end{equation}
An obvious solution for $\tau$-independent expectation values
$<\varphi^p(\tau)>$ are probability distributions for which $a_n,a_{n'}$
are nonvanishing only for one given value $n=n'$. Such
distributions correspond precisely to the stationary pure states
in the associated quantum mechanical problem. Furthermore, the
normalization (\ref{2.10}) implies $\bar a_n a_n=1$ and therefore
\begin{equation}\label{44.7}
<\varphi^p(\tau)>=K^p_{nn}\end{equation}
If the above solution is the only solution, we arrive at a
simple, perhaps surprising conclusion: The mere requirement
of $\tau$-independent $<\varphi^p(\tau)>$ associates the allowed
probability distributions to stationary quantum states. Furthermore,
the allowed values of $<\varphi^p>$ are discrete, reflecting
the discreteness of the associate quantum mechanical problem!

It remains to be shown that no other solutions exist with
$\partial_\tau<\varphi^p(\tau)>=0$. If we require the $\tau$-independence
for $<\varphi^p(\tau)>$ for all $p$, this can be established
in a rather straightforward way. Assume $\bar a_n\not=0$ for some
$n$. If $a_m$ is nonvanishing for $m\not=n$, the sum (\ref{44.6})
contains a $\tau$-dependent contribution $a_na_m
K^p_{nm}\exp\{(E_n-E_m)\tau\}$ at least for
certain values of $p$. (There is no superselection rule for the unharmonic
oscillator. Therefore for any given pair $(n,m)$ the integral $K^p_{nm}$
does not vanish for all $p$.) The above contribution to $\partial_\tau<\varphi^p(\tau)>$ could only be cancelled by
a similar contribution with the same $\tau$-dependence, whereby
$\bar a_{n'}$ and $a_{m'}$ must be nonzero and $E_{n'}-E_{m'}=E_n-E_m$,
with $n'\not=n,\ m'\not=m$. The spectrum of the unharmonic oscillator
has no degeneracy -- there is only one state for every eigenvalue
$E_n$. The equality of two energy differences in the spectrum could
occur accidentally for particular values of $\lambda$. This is,
however, not a problem for our argument since a cancellation of the
term $\sim \exp\{(E_n-E_m)\tau\}$ would only be possible at
the expense of new terms with a nontrivial $\tau$-dependence, i.e.
$\sim \exp\{(E_n-E_{m'})\tau\}$ and $\exp(E_{n'}-E_m)\tau$. This
problem will
be continued if one tries further to cancel the new terms. Since
equal energy differences can only occur accidently, one rapidly
arrives at a contradiction since no pair with the required precise
energy difference will be available any more. (A similar argument
would not work so easily for the harmonic oscillator with its
equidistant spectrum.)

Finally, if we require $\tau$-dependence only for some values of $p$,
the above argument will work only if for every pair $(m,n)$ there
exists a nonvanishing $K^p_{mn}$. Otherwise it is easy to
construct a counterexample. Assume we consider only $p=2$ and there is,
perhaps accidentally, a pair $(m,n)$ with $K^2_{mn}=0$.
Then a state $|\psi\}\sim(a_m\chi_m+a_n\chi_n)$ and similar for
$\{\psi|$ leads also to $\tau$-independent values of
$<\varphi^2(\tau)>$. The discreteness of the allowed values of
$\tau$-independent $<\varphi^2>$ would be lost in this case. However,
if a sufficient number of expectation values $<\varphi^p(\tau)>$
are required to be $\tau$-independent simultaneously, this
possibility of accidentally vanishing matrix elements can be discarded.

To conclude, the condition of $\tau$-independence of a suitable
number of expectation values $<\varphi^p(\tau)>$
leaves only a discrete number
of possibilities to continue the local
probability distribution (\ref{1.1}) outside
the local region. Correspondingly, the allowed $\tau$-independent
values $<\varphi^p>$ are also discrete. They correspond to the
stationary wave functions of the quantum mechanical unharmonic oscillator.
Since, in addition, $|\psi(\varphi)\}$ is real and positive one concludes
that excited states with zeros of $|\psi(\varphi)\}$ cannot be described in
our setting. The ground state solution can be realized by extending the
translation symmetry of $p_0$ also to $p_>$ and $p_<$, but also by all other
$p_>$ and $p_<$ which lead to the same state vectors $|\psi\}$ and $\{\psi|$.
We have seen that the
simple question of $\tau$-independence of expectation values has
led us directly to the appearance of quantum mechanical concepts
like states and operators in our classical statistical setting. In the
following we will exploit these quantum mechanical concepts to
describe also $\tau$-dependent expectation values. This will give
us a deeper insight how various quantum mechanical structures are
rooted in our formulation of incomplete classical statistics.

\bigskip
\section { Schr\"odinger and Heisenberg operators}
\setcounter{equation}{0}
\medskip
In this section we want to exploit further the mapping between
incomplete statistics and quantum mechanics for situations where
expectation values like $<\varphi(\tau)>$ may depend on $\tau$. A typical
question one may ask within incomplete classical statistics is the
following: Given a large set of measurements of observables with support
at a given value $\tau=0$, like $<\varphi^p(0)>, <(\partial_\tau\varphi(0))
^{p'}>$, etc., what can one predict for the expectation values of similar
observables at some other location $\tau\not=0$? It is obvious
that the evolution operator $\hat U$ is the appropriate tool to tackle
this type of questions.

The existence of the inverse evolution operator allows us to associate
to an observable $A(\tau)$ the operator $\hat A_S(\tau)$ in the
Schr\"odinger representation (cf. eq. (\ref{3.15}) \begin{equation}\label{4.1} \hat A_S(\tau)=\hat
U(\tau,\tau+\frac{\delta}{2})\hat A_\delta (\tau)\hat
U(\tau-\frac{\delta}{2},\tau)\end{equation}
The expectation value of the
observable $A$ can be expressed by the expectation value of the
operator $\hat A_S$ in a way analogous to quantum mechanics
\begin{equation}\label{4.2} <A(\tau)>=\{\psi(\tau)|\hat A_S(\tau)
|\psi(\tau)\} ={\rm
Tr}\rho(\tau)\hat A_S(\tau)\end{equation}
For the second identity we have
introduced the ``density matrix''
\begin{eqnarray}\label{4.3}
\rho(\varphi_1,\varphi_2,\tau)&=&|\psi(\varphi_1,\tau\}\{\psi(\varphi
_2,\tau)|=\int D\varphi_{(\tau'\not=\tau)}e^{-S(\varphi_1,\varphi_2)}
\nonumber\\ {\rm Tr}\rho(\tau)&=&1\end{eqnarray}
where $S(\varphi_1,\varphi_2)$
obtains from $S$ by replacing $\varphi(\tau)\to\varphi_1$ for all
``kinetic'' terms involving $\varphi(\tau'<\tau)$ and
$\varphi(\tau)\to\varphi_2$ for those involving $\varphi(\tau'>\tau)$,
whereas for potential terms $e^{-\epsilon V(\varphi(\tau))}\to
e^{-\frac{\epsilon}{2}(V(\varphi_1)+V(\varphi_2))}$.

In order to make the transition to the Heisenberg picture, we
may select a reference point $\tau=0$ and define
\begin{equation}\label{4.4} \hat
U(\tau)\equiv \hat U(\tau,0)\ ,\ \rho\equiv\rho(\tau=0)\ ,\
\rho(\tau)=\hat U(\tau)\rho\hat U^{-1}(\tau)\end{equation}
This specifies the
Heisenberg picture for the $\tau$-dependent operators
\begin{eqnarray}\label{4.5}
\hat A_H(\tau)&=&\hat U^{-1}(\tau)\hat A_S(\tau)\hat
U(\tau)\nonumber\\ <A(\tau)>&=&{\rm Tr}\rho \hat A_H(\tau)\end{eqnarray}
We note
that for two local observables $A_1, A_2$ the linear combinations
$A=\alpha_1A_1+\alpha_2A_2$ are also local observables. The associated
operators obey the same linear relations $\hat A=\alpha_1\hat
A_1+\alpha_2\hat A_2$, where $\hat A$ stands for $\hat A_\delta,\hat
A_S$ or $\hat A_H$. The relation (\ref{4.5}) is the appropriate
formula to answer the above question. One may use the set of measurements
of expectation values at $\tau=0$ to gather information about $\rho$.
Once $\rho$ is determined with sufficient accuracy, the expectation
values $<A(\tau)>$ can be computed. Of course, this needs a computation of
the explicit form of the Heisenberg operator $\hat A_H(\tau)$.

It is instructive to observe that some simple local observables have an
essentially $\tau$-independent operator representation in the
Schr\"odinger picture. This is easily seen for observables $A(\tau)$
which depend only on the variable $\varphi(\tau)$. The mapping reads
\begin{equation}\label{4.6}
A(\tau)=f(\varphi(\tau))\to\hat A_S(\tau)=f(\hat Q)\end{equation} For the
derivative operator $(\partial_\tau\varphi(\tau))^2$ a simple
representation obtains in the limit $\epsilon\to 0$
\begin{eqnarray}\label{4.7}
A(\tau)&=&(\partial_\tau\varphi(\tau))^2\to\nonumber\\ \hat
A_S(\tau)&=&\frac{1}{\epsilon}\left(\frac{1}{Z(\tau+\epsilon)+Z(\tau)}
+\frac{1}{Z(\tau-\epsilon)+Z(\tau)}\right)\nonumber\\ &&-\hat
P^2\left(\frac{2}{(Z(\tau+\epsilon)+Z(\tau))^2}
+\frac{2}{(Z(\tau-\epsilon)+Z(\tau))^2}\right)\nonumber\\
&&-2\Big\{\frac{2}{(Z(\tau+\epsilon)+Z(\tau))^2}e^{\frac{\epsilon}{2}
V(\hat Q,\tau)} \left[\hat P^2,e^{-\frac{\epsilon}{2}V(\hat
Q,\tau)}\right] \nonumber\\
&&-\frac{2}{(Z(\tau-\epsilon)+Z(\tau))^2}\left[\hat
P^2,e^{-\frac{\epsilon}{2}V(\hat Q,\tau)}
\right]e^{\frac{\epsilon}{2}V(\hat Q,\tau)}\Big\}\end{eqnarray}
The last term is
$\sim\epsilon^2$ and for constant $Z$ and $\epsilon\to 0$ one has
\begin{equation}\label{4.8}
(\partial_\tau\varphi(\tau))^2\to 1/(\epsilon Z)-\hat
P^2/Z^2\end{equation}
We observe that the r.h.s. of eq. (\ref{4.8}) differs from
$-\hat P^2/Z^2$ by a constant which diverges for $\epsilon\to0$. This
reconciles the positivity of $(\partial_\tau\varphi)^2$ with positive
\footnote{Note that $\hat{P}^2$ is not {\em necessarily} positive as long as
$|\psi(\tau)\}$ and $\{\psi(\tau)|$ are unrelated.} expectation values of $\hat P^2$.

For the transition to the Heisenberg picture the explicit definitions
(\ref{3.3}) and (\ref{3.10}) yield
\begin{equation}\label{4.9}
\hat Q\hat
U(\tau)=\varphi(\tau)\hat U(\tau)\ , \ \hat U(\tau)\hat
Q=\varphi(0)\hat U(\tau)\end{equation}
and  we note that the evolution operator
does not commute with $\hat Q$,
\begin{equation}\label{4.10} [\hat Q,\hat
U(\tau)]=(\varphi(\tau)- \varphi(0))\hat
U(\tau)=(\varphi_2-\varphi_1)\hat U(\varphi_2, \varphi_1;\tau,0)\end{equation}
Observables depending only on one variable $\varphi(\tau)$ have the
Heisenberg representation (cf. eq(\ref{4.6}))
 \begin{equation}\label{4.11}
A(\tau)=f(\varphi(\tau))\to \hat A_H(\tau)=\hat U^{-1}(\tau) f(\hat
Q)\hat U(\tau)=f(\hat Q(\tau))\end{equation}
Here we have used the definition
\begin{equation}\label{4.12}
\hat Q(\tau)=\hat U^{-1}(\tau)\hat Q\hat U(\tau)\end{equation}
More generally, one finds for products of functions depending on the
variables $\varphi(\tau_1),\varphi(\tau_2)...\varphi(\tau _n)$ with
$\tau_1<\tau_2<...\tau_n$ the Heisenberg operator
\begin{eqnarray}\label{4.13}
A(\tau_1,...\tau_n)&=&f_1(\varphi(\tau_1)f_2(\varphi(\tau_2))...
f_n(\varphi(\tau_n))\longrightarrow\nonumber\\ \hat A_H(\tau)&=&\hat
U^{-1}(\tau_n)f_n(\hat Q)\hat U(\tau_n, \tau_{n-1})...\hat
U(\tau_3,\tau_2)f_2(\hat Q)\hat U(\tau_2,\tau_1) f_1(\hat Q)\hat
U(\tau_1)\nonumber\\ &=&f_n(\hat Q(\tau_n))...f_2(\hat
Q(\tau_2))f_1(\hat Q(\tau_1))\end{eqnarray}
This important relation follows
directly from the definitions (\ref{2.8}), (\ref{4.1}), (\ref{4.5}). We
observe that $\hat A_H$ depends on the variables $\tau_i$ which are
the arguments of $A$ but shows no dependence on the reference point
$\tau$. (Only $\hat A_\delta$ and $\hat A_S$ depend on $\tau$.)

We can
use eq. (\ref{4.13}) to find easily the Heisenberg operators for
observables involving ``derivatives'', e.g.
\begin{eqnarray}\label{4.14}
A&=&\tilde\partial_\tau\varphi(\tau_1)=\frac{1}{2\epsilon}
(\varphi(\tau_1+\epsilon)-\varphi(\tau_1-\epsilon))\nonumber\\
\hat
A_H&=&\frac{1}{2\epsilon}\{\hat U^{-1}(\tau_1+\epsilon)\hat Q\hat U(
\tau_1+\epsilon)-\hat U^{-1}(\tau_1-\epsilon)\hat Q\hat
U(\tau_1-\epsilon)\} \nonumber\\
&=&\frac{1}{2\epsilon}\hat
U^{-1}(\tau_1)\{e^{\epsilon\hat H(\tau_1+\frac{\epsilon}{2})}\hat Q
e^{-\epsilon\hat
H(\tau_1+\frac{\epsilon}{2})}-e^{-\epsilon\hat
H(\tau_1-\frac{\epsilon}{2})}
 \hat Q e^{\epsilon\hat
H(\tau_1-\frac{\epsilon}{2})}\} \hat U(\tau_1)\nonumber\\
&=&\hat
U^{-1}(\tau_1)[\hat H(\tau_1),\hat Q]\hat U(\tau_1)+O
(\epsilon)\nonumber\\ &=&\frac{1}{2Z(\tau_1)}\hat U^{-1}(\tau_1)[\hat
P^2,\hat Q]\hat U(\tau _1)+O(\epsilon)\nonumber\\
&=&-\frac{1}{Z(\tau_1)}\hat U^{-1}(\tau_1)\hat R\hat
U(\tau_1)+O(\epsilon)=-\frac{1}{Z(\tau_1)}\hat R(\tau_1)+O(\epsilon)\end{eqnarray}
where we have assumed that $\hat H$ is a
smooth function of $\tau$. Here $\hat R$ is defined by
\begin{equation}\label{4.15} \hat
R(\varphi_2,\varphi_1)=\delta(\varphi_2-\varphi_1)\frac{\partial}
{\partial\varphi_1} \ ,\  \hat R^2=-\hat P^2\end{equation}
and we use, similar to eq. (\ref{4.12}), the definitions
\begin{equation}\label{5.15A}
\hat R(\tau)=\hat U^{-1}(\tau)\hat R\hat U(\tau)\ ,\quad \hat P
^2(\tau)=\hat U^{-1}(\tau)\hat P^2\hat U(\tau)\end{equation}

Two different definitions of derivatives can lead to the same
operator $\hat A_H$. An example is the observable
\begin{equation}\label{4.15a}
A=\partial^>_\tau\varphi(\tau_1)=\frac{1}{\epsilon}(\varphi(\tau_1+\epsilon)
-\varphi(\tau_1))\end{equation}
Up to terms of order $\epsilon$ the associated Heisenberg operator
is again given by $\hat A_H=-Z(\tau_1)^{-1}
\hat U^{-1}(\tau_1)\hat R\hat U(\tau_1)$
and therefore the same as for $\tilde\partial_\tau\varphi(\tau_1)$
(eq. (\ref{4.14})). Applying the
same procedure to the squared derivative observable (\ref{2.3}) yields
\begin{eqnarray}\label{4.16}
A&=&(\partial_\tau\varphi)^2(\tau_1)\longrightarrow\nonumber\\
\hat A_H&=&\frac{1}{2\epsilon^2}\hat U^{-1} (\tau_1)\{2\epsilon[\hat
Q,[\hat H,\hat Q]]+2\epsilon^2 [\hat Q,\hat H][\hat Q,\hat H]\}\hat
U(\tau_1)\nonumber\\
&=&\frac{1}{\epsilon Z}-\frac{1}{Z^2}\hat
U^{-1}(\tau_1) \hat P^2\hat U(\tau_1)=\frac{1}{\epsilon Z}-
\frac{1}{Z^2}\hat P^2(\tau_1)\end{eqnarray}
where we have assumed for
simplicity a $\tau$-independent Hamiltonian $\hat H$. This is in
agreement with
the direct evaluation (\ref{4.8}). It is remarkable that this operator
differs from the square of the Heisenberg operator associated
to $\tilde\partial_\tau\varphi(\tau_1)$ by a constant which diverges for $\epsilon\to0$. Indeed, one finds
\begin{eqnarray}\label{5.17}
A&=&(\tilde\partial_\tau\varphi(\tau_1))^2\to\hat A_H=\frac{1}{2\epsilon Z}-
\frac{1}{Z^2}\hat P^2(\tau_1)\nonumber\\
A&=&(\partial^>_\tau\varphi(\tau_1))^2\to \hat A_H=\frac{1}{\epsilon Z}-\frac{1}{Z^2}\hat P^2(\tau_1)
\end{eqnarray}
Incidentally, these relations can be used in order to find
an observable whose Schr\"odinger operator is given by $\hat P^2$ and
Heisenberg operator by $\hat P^2(\tau)$:
\begin{eqnarray}\label{5.18AA}
A_{P^2}(\tau_1)&=&Z^2(\partial^>_\tau\varphi(\tau_1))^2-2Z^2(\tilde\partial_\tau
\varphi(\tau_1))^2\nonumber\\
&=&\frac{Z^2}{\epsilon^2}\{\varphi^2(\tau_1)+\varphi(\tau_1+\epsilon
)\varphi(\tau_1-\epsilon)-\varphi(\tau_1+\epsilon)\varphi(\tau_1)-
\varphi(\tau_1)\varphi(\tau_1-\epsilon)\}\nonumber\\
&&\to \hat A_H=\hat P^2(\tau_1)\end{eqnarray}

As a last instructive example for the association between observables
and Heisenberg operators we consider (for $\epsilon\to 0$)
\begin{eqnarray}\label{5.18A}
A&=&\partial^>_\tau\varphi(\tau_1)\varphi(\tau_2)\quad\longrightarrow
\nonumber\\
\hat A_H&=&\left\{\begin{array}{lll}
-Z^{-1}\hat R(\tau_1)\hat Q(\tau_2)&
{\rm for}& \tau_1\geq\tau_2+\epsilon\\
-Z^{-1}\hat Q(\tau_2)\hat R(\tau_1)&
{\rm for}& \tau_1\leq\tau_2-\epsilon\end{array}\right.\end{eqnarray}
The place were $\hat R$ and $\hat Q$ appear depends on the ordering
of $\tau_1$ and $\tau_2$. One obtains the same Heisenberg operator (\ref{5.18A}) if one replaces $\partial^>_\tau\varphi(\tau_1)$ by
$\tilde\partial_\tau\varphi(\tau_1)$. On the other hand, for equal
locations $\tau_2=\tau_1$ the two observables correspond to different
Heisenberg operators
\begin{eqnarray}\label{5.18B}
\varphi(\tau_1)\partial^>_\tau\varphi(\tau_1)&\to &-Z^{-1}\hat U^{-1}(\tau_1)\hat R\hat Q
\hat U(\tau_1)\nonumber\\
\varphi(\tau_1)\tilde\partial_\tau\varphi(\tau_1)&\to &-\frac{1}{2}
Z^{-1}\hat U^{-1}(\tau_1)(\hat R\hat Q
+\hat Q\hat R)\hat U(\tau_1)\nonumber\\
&=&-Z^{-1}\hat U^{-1}(\tau_1)\hat R\hat Q\hat U(\tau_1)+\frac{1}{2}
Z^{-1}\end{eqnarray}
One concludes that for arbitrary probability distributions $p_<$ and
$p_>$ (or states $\{\psi|$ and $|\psi\}$) the expectation values
$<\partial^>_\tau\varphi(\tau_1)\varphi(\tau_1)>$ and $<\tilde \partial
_\tau\varphi(\tau_1)\varphi(\tau_1)>$ differ.

Eqs. (\ref{5.17}) and (\ref{5.18B}) teach us that the product
of derivative observables with other observables can be
ambiguous in the sense that the associated operator and expectation
value depends very sensitively on the precise definition of the
derivative.
This ambiguity of
the derivative observables in the continuum limit is an unpleasant
feature for the formulation of correlation functions. In the next sections
we will see how this problem is connected with the concept of quantum
correlations. We will argue that the ambiguity in the classical
correlation may be the basic ingredient why a description of our
world in terms of quantum statistics is superior to the use of
classical correlation functions.

\bigskip
\section{Correlation functions in incomplete classical\protect\\ statistics}
\setcounter{equation}{0}

A basic concept for any statistical description are correlation functions
for a number of observables $A_1[\varphi], A_2[\varphi], ...$ In particular,
a two-point function is given by
the expectation value of an associative product of two observables
$A_1[\varphi]$ and $A_2[\varphi]$. For local observables $A_1,A_2$ the
product should again be a local observable which must be defined
uniquely in terms of the definitions of $A_1$ and $A_2$. This
requirement, however, does not fix the definition of the correlation
uniquely. The standard ``classical product'', i.e. the simple
multiplication of the functionals $A_1[\varphi]\cdot A_2[\varphi]$
(in the same sense as the ``pointwise'' multiplication of functions)
fulfills the general requirements\footnote{This holds provided that
the product results in a meaningful observable with finite expectation
value.} for a correlation function. Other
definitions can be conceived as well. In this section we will introduce
a quantum correlation which equals the classical (``pointwise'')
correlation only for $\tau$-ordered non-overlapping observables.
In contrast, for two local observables with overlapping support
we will find important differences between the quantum and classical
correlation. In particular, we will discover the
effects of the non-commutativity characteristic for quantum mechanics.

Incomplete statistics draws our attention to
an important issue in the formulation of meaningful correlation functions.
 Consider the two versions
of the derivative observable $\tilde\partial_\tau\varphi$ and
$\partial^>_\tau\varphi$ defined by eqs. (\ref{4.14}) and
(\ref{4.15}), respectively. In the continuum limit $(\epsilon\to 0)$ they
are represented by the same operator $\hat A_H$. In consequence,
both definitions lead to the same expectation value for any state
$|\psi\},\{\psi|$.  The two
versions of derivative observables cannot be distinguished
by any measurement and should
therefore be identified. On the other hand, the classical products
$\tilde\partial_\tau\varphi(\tau_1)\cdot\tilde\partial_\tau
\varphi(\tau_2)$ and $\partial^>_\tau\varphi(\tau_1)\cdot\partial
_\tau^>\varphi(\tau_2)$ are represented by different operators
for $\tau_1=\tau_2$,
as can be seen from eq. (\ref{5.17}). (See
also eq. (\ref{5.18B}) for a similar argument involving
$\partial_\tau\varphi(\tau_1)\cdot\varphi(\tau_2)$.)
This means that  the two versions of derivative
observables lead to different classical correlation functions! Obviously,
this situation is unsatisfactory since for $\epsilon\to 0$
no difference between the two versions could  be ``measured''
for the observables themselves. We find this disease
unacceptable for a meaningful correlation and require
as a criterion for a meaningful correlation
function  that two observables which have the same
expectation values for all (arbitrary) probability
distributions should also have identical correlation functions.
We have shown that two observables
which are represented by the same Heisenberg operator have indeed
the same expectation values for all possible probability
distributions and should therefore be considered as equivalent.
They should therefore
lead to indistinguishable correlation functions.

As we have already established, the two derivative observables
$A_1=\tilde\partial_\tau
\varphi(\tau)$ and $A_2=\partial^>_\tau\varphi(\tau)$ are
indistinguishable in the continuum limit, whereas their classical correlations are not.
We may therefore conclude \footnote{
We will discuss in the next section to what extent this problem may
be cured by a restriction to better ``smoothened'' observables.}
that the classical correlation $A_1\cdot A_2$ is not a meaningful correlation function.
In this section we propose the use of a different correlation
based on a quantum product $A_1\circ A_2$. By construction,
this correlation will always obey our criterion of ``robustness''
with respect to the precise choice of the observables. It should
therefore be considered as an interesting alternative to
the classical correlation. At this place we only note that the
``robustness problem'' is not necessarily connected to
the continuum limit. The mismatch between
indistinguishable observables and distinguishable ``classical'' correlations
can appear quite generally also for $\epsilon>0$.

Our formulation of a quantum correlation will be based on the concepts
of  equivalent observables and products defined for equivalence
classes. In fact, the
mapping $A(\tau)\to\hat A_H(\tau)$ is not necessarily
invertible on the space
of all observables $A(\tau)$. This follows from the simple observation
that already the map (\ref{2.8}) contains integrations.
Two different integrands (observables)
could lead to the same value of the integral (operator)
for arbitrary  fixed boundary
values $\varphi(\tau-\frac{\delta}{2}),\ \varphi(\tau+\frac{
\delta}{2})$. It is
therefore possible that two different observables $A_a
(\tau)$ and $A_b(\tau)$ can be mapped into the same
Heisenberg operator $\hat
A_H(\tau)$. Since the expectation values can be computed from $\hat
A_H(\tau)$ and $\rho$ only, no distinction between $<A_a>$ and $<A_b>$
can then be made for arbitrary $\rho$. All local observables $A(\tau)$
which correspond to the same operator $\hat A_H(\tau)$ are
equivalent.

We are interested in structures that only
depend on the equivalence classes of observables. Addition of two
observables and multiplication with a scalar can simply be carried
over to the operators. This is not necessarily the case, however, for the
(pointwise) multiplication of two observables.  If $A_a{(\tau)}$ and
$A_b{(\tau)}$ are both mapped into $\hat A_H(\tau)$ and a third observable
$B(\tau)$ corresponds to $\hat B_H(\tau)$, the
products $A_a\cdot B$ and $A_b \cdot B$ may nevertheless be
represented by different operators (cf. eq. (\ref{5.18AA}), (\ref{5.18B})).
It is then easy to construct states
where $<A_aB> \not=$$<A_bB>$ and the pointwise product does not
depend only on the equivalence class.

On the other
hand, the (matrix) product of two operators $\hat A_H \hat B_H$
obviously refers only to the equivalence class. It can be implemented
on the level of observables by defining a unique ``standard
representative'' of the equivalence class as
\begin{equation}\label{5.1} \bar
A[\varphi,\tau]=F[\hat A_H(\tau)]\end{equation}
Using the mapping $A[\tau]\to\hat
A_H(\tau)$ (\ref{2.8}), (\ref{4.1}), (\ref{4.5}), we define
the quantum product of
two observables as
\begin{equation}\label{5.2}
A(\varphi,\tau)\circ
B(\varphi,\tau)=F[\hat A_H(\tau)\hat B_H(\tau)] \equiv(A\circ
B)[\varphi,\tau]\end{equation}
This product is associative, but not commutative. (By
definition, the operator associated to the observable $(A\circ
B)(\varphi,\tau)$ is $\hat A_H(\tau)\hat B_H(\tau)$ and the product
$A\circ B$ is isomorphic to the ``matrix multiplication'' $\hat A\hat
B$ if restricted to the subspace of operators $\bar A=F[\hat A], \bar
B=F[\hat B]$.)
The correlations (e.g. expectation values of products of
observables) formed with the product $\circ$ reflect the
non-commutative structure of quantum mechanics.
This justifies the name ``quantum correlations''.
Nevertheless, we emphasize that the
``quantum product'' $\circ$ can also be viewed as
just a particular structure among
``classical observables''.

The definition of the quantum product is unique on the level of operators.
On the level of
the classical observables, it is, however, not yet fixed uniquely
by eq. (\ref{5.2}). The precise definition
obviously depends on the choice of a
standard representation $F[\hat A_H(\tau)]$ for the equivalence class of
observables represented by $\hat{A}_H$. We will  choose a linear
map $F[\alpha_1\hat A_{H,1}+\alpha_2\hat A_{H,2}] =\alpha_1F[\hat
A_{H,1}]+\alpha_2 F[\hat A_{H,2}]$ with the property that it inverses
the relation (\ref{4.13}). For ``time-ordered''
$\tau_1<\tau_2<...\tau_n$ the map $F$ should then obey
\begin{equation}\label{5.3}
F[f_n(\hat Q(\tau_n))...f_2(\hat Q(\tau_2))f_1(\hat Q(\tau_1))]
=f_1(\varphi(\tau_1))f_2(\varphi(\tau_2))...f_n(\varphi(\tau_n)).\end{equation}
It is easy to
see how this choice exhibits directly the noncommutative
property of the quantum
product between two observables. As an example let us consider the two
observables $\varphi(\tau_1)$ and $\varphi(\tau_2)$ with
$\tau_1<\tau_2$. The quantum product or quantum correlation
depends on the ordering
\begin{eqnarray}\label{5.4}
\varphi(\tau_2)\circ\varphi(\tau_1)&=&\varphi(\tau_2)\varphi(\tau_1)
\nonumber\\
\varphi(\tau_1)\circ\varphi(\tau_2)&=&\varphi(\tau_2)
\varphi(\tau_1)+F[[\hat Q(\tau_1),\hat Q(\tau_2)]]\end{eqnarray}
The
noncommutative property of the quantum product for these operators is
directly related to the commutator
\begin{eqnarray}\label{5.5} [\hat Q(\tau_1),\
\hat Q(\tau_2)]&=&\hat U^{-1}(\tau_1)\hat Q\hat   U(\tau_1,\tau_2)\hat
Q \hat U(\tau_2)\nonumber\\
&&-\hat U^{-1}(\tau_2)\hat Q\hat
U(\tau_2,\tau_1)\hat Q\hat U(\tau_1)\end{eqnarray}
Only for time-ordered arguments the quantum correlation coincides
with the classical correlation. The nonvanishing commutator $\varphi(\tau_1)\circ\varphi(\tau_2)-\varphi(\tau_2)\circ\varphi(\tau_1)$
could presumably play a crucial role in the
discussion of Bell's inequalities
\cite{Bell} once our system is extended to a setting with more than
one particle. \footnote{Note that Bell's inequalities are based on the
definition of the correlation function by ``pointwise multiplication''.}

The map $F$ can easily be extended to operators involving
derivatives of $\varphi$. We concentrate here for simplicity on
a translation invariant probability distribution in the local
region with constant $Z(\tau)=Z$.
The mappings (with $\tau_2\geq \tau_1+\epsilon)$
\begin{eqnarray}\label{6.6}
F(\hat R(\tau))&=&-Z\partial^>_\tau\varphi(\tau)\nonumber\\
F(\hat R(\tau)\hat Q(\tau))&=&-Z\varphi(\tau)\partial^>_\tau\varphi
(\tau)\nonumber\\
F(\hat R(\tau_2)\hat R(\tau_1))&=&Z^2\partial^>_\tau\varphi(\tau_2)\partial^>_\tau\varphi(
\tau_1)\end{eqnarray}
are compatible with eq. (\ref{5.3}). This can be seen by noting that
the $\tau$-evolution of $\hat Q(\tau)$ according to eq. (\ref{4.12})
implies for $\epsilon\to 0$ the simple relation
\begin{equation}\label{6.7}
\partial_\tau\hat Q(\tau)=[\hat H,\hat Q(\tau)]=-Z^{-1}\hat R(\tau)\end{equation}
A similar construction (note $[\hat Q(\tau+\epsilon), \hat Q(\tau)]
=-\epsilon/Z)$ leads to
\begin{equation}\label{6.8}
F(\hat R^2(\tau))=Z^2(\partial_\tau^>\varphi(\tau))^2-Z/\epsilon\end{equation}
and we infer that the quantum product of derivative observables at
equal sites differs from the pointwise product
\begin{equation}\label{6.9}
\partial^>_\tau\varphi(\tau)\circ\partial^>_\tau\varphi(\tau)=(\partial
^>_\tau\varphi(\tau))^2-1/(\epsilon Z)\end{equation}

From the relations (\ref{5.4}) and (\ref{6.9}) it has become clear
that the difference between the quantum product and the ``pointwise''
classical product of two observables is related to their $\tau$-ordering
and ``overlap''. Let us define that two observables $A_1[\varphi]$
and $A_2[\varphi]$ overlap if they depend on variables $\varphi(\tau)$
lying in two overlapping $\tau$-ranges ${\cal R}_1$ and ${\cal R}_2$.
(Here two
ranges do not overlap if all $\tau$ in ${\cal R}_1$ obey $\tau\leq \tau_0$ whereas
for ${\cal R}_2$ one has $\tau\geq \tau_0$, or vice versa.
This implies that non-overlapping observables can depend on at
most one common variable $\varphi(\tau_0)$.) With this definition
the quantum product is equal to the classical product if the observables
do not overlap and are $\tau$-ordered (in the sense that larger $\tau$
are on the left side).

In conclusion, we have established a one-to-one correspondence between
classical correlations $\varphi(\tau_2)\varphi(\tau_1)$ and the
product of Heisenberg operators $\hat Q(\tau_2)\hat Q(\tau_1)$ provided
that the $\tau$-ordering $\tau_2\geq\tau_1$ is respected. This extends
to observables that can be written as sums or integrals over $\varphi(\tau)$
(as, for example, derivative observables) provided the $\tau$-ordering
and non-overlapping properties are respected. For well separated
observables no distinction between a quantum and classical $\tau$-ordered
correlation function would be needed. In particular, this holds also
for ``smoothened'' observables $A_i$ that involve (weighted) averages
over $\varphi(\tau)$ in a range ${\cal R}_i$ around $\tau_i$. Decreasing the
distance between $\tau_2$ and $\tau_1$, the new features of the quantum
product $A_1(\tau_2)\circ A_1(\tau_1)$ show up only once the distance
becomes small enough so that the two ranges ${\cal R}_1$ and ${\cal R}_2$
start to
overlap. In an extreme form the difference between quantum and classical
correlations becomes apparent for derivative observables at the same
location. Quite generally, the difference between the quantum and
classical product is seen most easily on the level of the
associated operators
\begin{eqnarray}\label{6.10}
A_1\circ A_2&\to& \hat A_1\hat A_2\nonumber\\
A_1\cdot A_2&\to& T(\hat A_1\hat A_2)\end{eqnarray}
Here $T$ denotes the operation of $\tau$-ordering. The $\tau$-ordered
operator product is commutative $T(\hat A_1\hat A_2)=T(\hat A_2\hat A_1)$
and associative $T(T(\hat A_1\hat A_2)\hat A_3)=
T(\hat A_1 T(\hat A_2\hat A_3))\equiv T(\hat A_1\hat A_2\hat A_3)$.
As we have seen in the discussion of the derivative observables,
it lacks, however, the general property of robustness with respect
to the precise definition of the observables. This contrasts with the
non-commutative product $\hat A_1\hat A_2$.

\section{Complete classical statistics, irrelevant and inaccessible
information}
\setcounter{equation}{0}
Our discussion of incomplete classical statistics may perhaps have led to
the impression that the quantum mechanical properties are somehow related
to the missing information. This is by no means the case! In fact, our
investigation of the consequences of incomplete information about
the probability distribution was useful in order to focus the
attention on the question which information is really necessary to
compute the expectation values of local observables. We can now turn
back to standard ``complete'' classical statistics where a given
probability distribution $p[\varphi(\tau)]$ is assumed to be known.
We concentrate here on a general class of probability distributions
which can be factorized in the form $p=p_>p_0p_<$ according
to eq. (\ref{1.1}) -- it may be called ``factorizable'' or
``$F$-statistics''. For example, all systems which have only
local and next-neighbor interactions are of this form. Within
$F$-statistics the states remain defined according to eq. (\ref{2.7}).

We emphasize that any additional information contained in
$p[\varphi]$ which goes beyond the local distribution $p_0[\varphi]$ and the
states $|\psi\}$ and $\{\psi|$ does not change a iota in
the expectation values of local observables and their correlations!
The additional information is simply {\em irrelevant} for the
computation of local expectation values. A given probability
distribution specifies $p_<$ and $p_>$ uniquely. This determines
$|\psi\}$ and $\{\psi|$ and we can then
continue with the preceding discussion in order to calculate the
expectation values of local observables. The precise form of
$p_<$ and $p_>$ which has led to the given states plays no role
in this computation.

Since all information contained in $p_<$ and $p_>$ beyond the states
$|\psi\}$ and $\{\psi|$ is irrelevant for local expectation values,
it is also {\em inaccessible} by any local measurements. In fact,
even the most precise measurements of expectation values and
correlation functions for arbitrarily many local observables
could at best lead to a reconstruction of the states $|\psi\}$ and $\{
\psi|$. This sheds new light on the notion of ``incompleteness''
of the statistical information discussed in this note. In fact,
within $F$-statistics the ``incomplete'' information contained in
the states $|\psi\}$ and $\{\psi|$ constitutes the most complete
information that can possibly be achieved by local measurements! Since
any real measurement is local in time and space all assumptions
about information beyond the states concern irrelevant and
inaccessible information and cannot be verified by observation!

Particularly interesting in the spirit of the completeness
question asked by Einstein is the following situation: Suppose first
that an observer has no information about the ``exterior region'' beyond
the one contained in the state vectors. Nevertheless, his information
is sufficient in order to compute the expectation values and correlations
of all local observables. What happens, however, if
additional information about the probability distribution in the
exterior region becomes available?
Our observer could now perhaps compute additional expectation values for
observables with support outside the local region. As far as the local
observables are concerned, however, the additional information is completely
irrelevant. In this sense quantum mechanics is complete as far
as local observables are concerned.

These simple remarks have striking consequences: all the quantum
mechanical features found in our discussion of incomplete classical
statistics are genuine properties of {\em all} classical
statistical systems whose probability distribution can be factorized.
This concerns, in particular, the expectation values of the observables
$\varphi^p(\tau)$. For an arbitrary $\tau$-translation invariant and
factorizable probability distribution all $<\varphi^p(\tau)>$ will
be independent of $\tau$ unless the translation symmetry is spontaneously
broken. We infer that only the discrete stationary values of $<\hat Q^p>$ of
the quantum-mechanical system with the appropriate $\hat H$ are possible
solutions. We see how ``quantum mechanical''
information can be used directly for a discussion of classical
statistical properties in a wide class of systems.

Another possibility concerns the use of the uncertainty relation which
is also present in classical $F$-statistics. Assuming for simplicity
a discrete symmetry $\varphi\to-\varphi$ with $<\varphi>=0$ this
yields the inequality (cf. the definition (\ref{5.18AA}))
\begin{equation}\label{N7.1}
<\varphi^2(\tau)><A_{P^2}(\tau)>=<\hat Q^2><\hat P^2>\geq 1/4\end{equation}
(Throughout this paper we use conventions for the units of $\varphi$
and $\tau$ where $\hbar\equiv1$.)
It is not obvious how such an inequality for the derivative-like
observable $A_{P^2}$ (\ref{5.18AA}) would have been found without using
the simple commutator relations of quantum-mechanical operators.
The simple fact that quantum-mechanical information can be used in
practice to establish properties of expectation values in a standard
classical statistical system demonstrates in a simple way that
quantum-mechanical features are indeed genuine properties of
classical statistical systems.

The quantum correlation between two observables can be formulated
in standard classical statistics as well. Together with the
``classical'' correlation function we have thus more than one
candidate for the definition of a correlation function in a given
classical statistical system\footnote{Actually, the same situation
would arise had we started from a quantum-mechanical system: one has
to decide if the noncommuting operator product $\hat A\hat B$ or
rather the commuting $\tau$-ordered product $T(\hat A\hat B)$ should
be used for the definition of the correlation.}.
The final decision which correlation is relevant in practice depends
on the question which type of correlation can typically be measured
by experiment. We believe that a certain ``robustness'' of the
correlation with respect to the precise choice of the observable
plays an important role in this respect.

We have argued in the preceding section that the classical correlation
may be disfavored by the lack of robustness for local correlations
of local derivative operators. The observed disease of the classical
correlation gets weakened, however, once we consider the average
of a correlation $<A(\tau_1)B(\tau_2)>$ over a certain range
in $|\tau_2-\tau_1|$. With the mappings from observables to Heisenberg
operators (for $\epsilon\to0,\ Z(\tau)=Z)$
\begin{eqnarray}\label{N7.2}
&&\partial^>_\tau\varphi(\tau+\epsilon)\partial^>_\tau\varphi(\tau)\to \frac
{1}{Z^2}\hat R(\tau+\epsilon)\hat R(\tau)=-\frac{1}{Z^2}\hat P^2(\tau)
\nonumber\\
&&\tilde\partial_\tau\varphi(\tau+\epsilon)\tilde\partial_\tau\varphi
(\tau)=\frac{1}{4\epsilon Z}-\frac{1}{Z^2}\hat P^2\end{eqnarray}
we see that the difference between the two versions of derivative
operators vanishes for an average of classical correlations which is defined by
\begin{equation}\label{N7.3}
\frac{1}{3}\sum^1_{m=-1}<\partial_\tau\varphi(\tau+m\epsilon)
\partial_\tau\varphi(\tau)>\end{equation}
Similarly, for $\Delta=(2n+1)\epsilon$ we may define averaged
observables
\begin{equation}\label{N7.4}
A^\Delta(\tau)=\frac{1}{2n+1}\sum^n_{m=-n}A(\tau+\epsilon n)\end{equation}
and we find for the difference of the square of the two versions
of derivative operators
\begin{equation}\label{N7.5}
(\partial^>_\tau\varphi)^\Delta(\tau)(\partial_\tau^>\varphi)^\Delta
(\tau)-(\tilde\partial_\tau \varphi)^\Delta(\tau)(\tilde\partial_\tau\varphi)^\Delta(\tau)
\to\frac{\epsilon}{2\Delta^2}\end{equation}
For fixed $\Delta$ this vanishes for $\epsilon\to0$ and one may
incorrectly conclude that the problem of the lack of robustness
for classical correlation functions disappears for smoothened or
averaged observables.

That this is not the case can be realized by the comparison of two
other smoothened derivative observables
\begin{eqnarray}\label{N7.6}
\partial^{\Delta_>}_\tau\varphi(\tau)&=&
\frac{\varphi(\tau+\Delta)-\varphi(\tau)}{\Delta},\nonumber\\
\tilde\partial^{\Delta}_\tau\varphi(\tau)&=&
\frac{\varphi(\tau+\Delta)-\varphi(\tau-\Delta)}{2\Delta}\end{eqnarray}
It is straightforward to verify that both $(\partial^>_\tau\varphi)^\Delta$
and $(\tilde\partial_\tau\varphi)^\Delta$ actually correspond
to the observable $\tilde\partial_\tau^\Delta\varphi$ up to
corrections $O(\epsilon)$ whereas for a discussion of the robustness
problem on the scale $\Delta$ we should compare two versions
of derivative observables that may differ in order
$O(\Delta)$ as the two versions in eq. (\ref{N7.6}):
\begin{eqnarray}
\partial^{\Delta_>}_\tau\varphi(\tau)&\to&-\frac{1}{Z}\hat R(\tau)+O(\Delta)
\nonumber\\
\tilde\partial^{\Delta}_\tau\varphi(\tau)&\to&-\frac{1}{Z}\hat R(\tau)+O(\Delta)\end{eqnarray}
Nevertheless, the classical correlation function differs for the
two choices by a term that does not vanish for $\Delta\to0$: One finds,
for $|\Delta<\hat H>|\ll 1$ and $\tau_2\geq \tau_1$:
\begin{eqnarray}\label{N7.8}
&&\partial_\tau^{\Delta_>}\varphi(\tau_2)\partial_\tau^{\Delta_>}
\varphi(\tau_1)-\tilde\partial_\tau^\Delta(\tau_2)
\tilde\partial_\tau^\Delta\varphi(\tau_1)\nonumber\\
&&\nonumber\\
&&\to\frac{\Delta-(\tau_2-\tau_1)}{\Delta^2Z}\theta(\Delta-(\tau_2-\tau_1))-
\frac{\Delta-\frac{1}{2}(\tau_2-\tau_1)}{2\Delta^2Z}\theta
(2\Delta-(\tau_2-\tau_1))\end{eqnarray}
In contrast to eq. (\ref{N7.5}) this difference survives in the
continuum limit $\epsilon\to 0$. Furthermore, for $\tau_2=\tau_1$ it
``diverges'' again for small $\Delta$ as $1/(2\Delta Z)$.
We conclude that the lack of
robustness of the classical correlation remains present for
averaged observables as well!

To be more precise, we will formulate the following ``robustness
criterion'' for meaningful correlation functions: if two smoothened
observables with support in a $\tau$-range of the order $O(\Delta)$
have equal expectation values for all probability distributions
up to corrections $\sim O(\Delta)$, then the respective correlation
functions should also become identical up to corrections $\sim O(\Delta)$.
(Here it is understood that $|\Delta<\hat H>|\ll1$.) Form eq.
(\ref{N7.8}) we learn that the classical correlation does not meet
this robustness criterion. For example, for $\tau_2$ near $\tau_1$ the
difference between the correlation for the two versions of the
derivative operator ``diverges'' $\sim \Delta^{-1}$. In fact, this disease
extends to a whole range of $(\tau_2-\tau_1)\sim O(\Delta)$.
On the other hand, the quantum correlation obeys the robustness
criterion by construction.

This discussion opens an interesting perspective: The difference
between classical and quantum statistics seems to be  a
question of the appropriate definition of the correlation function.
Simple arguments of robustness favor the choice of the quantum correlation!
In a sense, the successful description of nature by quantum-mechanical
operators and their products
gives an ``experimental indication'' that quantum correlations should be
used!

\bigskip
\section{ Superposition of states and density matrices}
\setcounter{equation}{0}
\medskip
A fundamental principle of quantum mechanics is the superposition of
states. It states that for two quantum states $|\psi^{(1)}\rangle$ and
$|\psi^{(2)}\rangle$ the linear superposition
$|\psi\rangle=\alpha_1|\psi^{(1)}\rangle +\alpha_2|\psi^{(2)}\rangle$
is again a possible quantum state. (Proper normalization is assumed.)
Furthermore, for two density matrices $\rho^{(1)},\rho^{(2)}$ the
linear combination $\rho=w_1\rho^{(1)}+w_2\rho^{(2)}, \ w_1+w_2=1$, is
again a possible density matrix if it obeys the appropriate positivity
conditions.  We will show that these properties arise naturally in our
context of incomplete classical statistics (or, more generally, F-statistics),
at least for a restricted choice of $\alpha_1,\alpha_2$.

Let us consider two probability distributions
$p^{(1)}[\varphi]=\exp(-S^{(1)}[\varphi],p^{(2)}
[\varphi]=\exp(-S^{(2)}[\varphi])$
which  differ only outside the local range $-\bar\tau<\tau<\bar\tau$,
i.e.
\begin{eqnarray}\label{10.1}
S^{(i)}[\varphi]&=&S_0^{(i)}[\varphi]+S_>^{(i)}[\varphi]+S^{(i)}_<[\varphi],
\nonumber\\
S_0^{(1)}[\varphi]&=&S_0^{(2)}[\varphi]=S_0[\varphi]
=\int^{\bar\tau}_{-\bar\tau}
d\tau'{\cal L}(\tau'),\nonumber\\
S_>^{(1)}[\varphi]&\not=&S_>^{(2)}[\varphi]\ ,\  S_<^{(1)}[\varphi]\not=
S^{(2)}_<[\varphi]\end{eqnarray}
Within the local range the two probability distributions correspond to
the same dynamics
\begin{equation}\label{10.2}
\hat H^{(1)}(\tau)=\hat H^{(2)}(\tau),\ \hat U^{(1)}(\tau)=\hat
U^{(2)}(\tau)\end{equation}
whereas the states $|\psi^{(1)}\}$ and $|\psi^{(2)}\}$ differ. We will
assume that both $S^{(1)}$ and $S^{(2)}$ are invariant under the
reflection symmetry. As far as local observables are concerned the
corresponding operators are the same for both situations. The only
difference in the expectation values of observables between the two
ensembles  can be traced back to different state vectors at some
reference point $\tau_0$ or time $t_0$. This setting reflects
precisely the situation for two different quantum mechanical states
$|\psi_1(t_0)\rangle\not=|\psi_2(t_0)\rangle$.

A superposition state corresponds to a new probability distribution
given by $S[\varphi]=S_0[\varphi]+S_>[\varphi]+S_<[\varphi]$ with
\begin{equation}\label{10.3}
\exp(-S_>[\varphi])=\alpha_1\exp(-S_>^{(1)}[\varphi])+\alpha_2\exp
(-S^{(2)}_>[\varphi])\end{equation}
Here the real coefficients $\alpha_1$ and $\alpha_2$ have to obey the
condition that $\exp(-S[\varphi])$ is positive  semidefinite for all
$\varphi$ and properly normalized.
 The definition (\ref{2.7}) of the states as functional
integrals implies directly
\begin{equation}\label{10.4}
|\psi(\bar\tau)\}=\alpha_1|\psi^{(1)}(\bar\tau)\}+\alpha_2|\psi^{(2)}
(\bar\tau)\}\end{equation}
We observe that the superposition is compatible with the evolution
such that for all $\tau$ in the interval  $-\bar\tau<\tau<\bar\tau$ one finds
$|\psi(\tau)\}=\alpha_1|\psi^{(1)}(\tau)\}+\alpha_2|\psi^{(2)}(\tau)\}$.
We conclude that quantum mechanical superposition arises directly
from the construction of the probability distribution (\ref{10.3}). In
particular, we emphasize  the appearance of interference in the
computation of expectation values of operators
\begin{eqnarray}\label{10.6}
\langle
A\rangle=&&\langle\psi|A^{(M)}|\psi\rangle=\alpha^2_1
\langle\psi^{(1)}|A^{(M)}|\psi^{(1)}\rangle+\alpha^2_2
\langle\psi^{(2)}|A^{(M)}|\psi^{(2)}\rangle
\nonumber\\
&&+\alpha_1\alpha_2(\langle\psi^{(1)}|A^{(M)}|\psi^{(2)}\rangle+\langle
\psi^{(2)}|A^{(M)}|\psi^{(1)}\rangle)
\end{eqnarray}
The interference terms $\sim\alpha_1\alpha_2$ are characteristic for
``quantum statistics''. In our approach they are connected to the
fact that the probability distribution corresponding  to (\ref{10.3})
cannot be written as a sum of two probability distributions! It should be
emphasized that $\alpha_1$ and $\alpha_2$ may have opposite sign such that
interference may indeed be destructive. Nevertheless, the condition stipulated
after eq. (\ref{10.3}) does not admit arbitrary superpositions. We note that the
linear structure of the Schr\"odinger equation and the quantum
mechanical time evolution arise directly from the basic
construction. Nonlinear generalizations of quantum mechanics would
have to modify this structure.

In distinction to the construction (\ref{10.3}) we can also consider
linear combinations of probability distributions
\begin{equation}\label{10.7}
p[\varphi]=\exp(-S[\varphi])=w_1\exp(-S^{(1)}[\varphi])+w_2\exp
(-S^{(2)}[\varphi])\end{equation}
This alternative construction does not lead to interference and
results in a linear combination of density matrices
(cf. eq. (\ref{4.2}))
\begin{equation}\label{10.8}
\rho=w_1\rho^{(1)}+w_2\rho^{(2)}\end{equation}
with
\begin{equation}\label{10.9}
\langle A\rangle=\Tr(\hat A\rho)=w_1\Tr(\hat A\rho^{(1)})+w_2\Tr(\hat
A\rho^{(2)})\end{equation}
The density matrix $\rho$ (\ref{10.8}) is not anymore the density matrix
corresponding to a pure quantum mechanical state. In particular, the
 ``pure state density matrices'' $\rho^{(1)},\rho^{(2)}$
obey
\begin{equation}\label{10.10}
\Tr(\rho^{(i)})^2=\Tr(\rho^{(i)})=1\end{equation}
whereas (for $w_i\not= 0$)
\begin{eqnarray}\label{10.11}
\Tr\rho^2&=&w_1^2\Tr(\rho^{(1)})^2+w^2_2
\Tr(\rho^{(2)})^2+2w_1w_2\Tr(\rho^{(1)}\rho^{(2)})\nonumber\\
&=&1+2w_1w_2(\Tr(\rho^{(1)}\rho^{(2)})-\Tr\rho^{(1)}\Tr\rho^{(2)})\not=1.\end{eqnarray}

We emphasize that the probability distribution (\ref{10.7}) cannot be
written in the product form (\ref{1.1}) any more. The linear
combination of density matrices therefore generalizes our
original concept of incomplete statistics. This contrasts to
the superposition of states (\ref{10.3}) which preserves the product
structure (\ref{1.1}). For an extended concept of incomplete statistics
with probability distributions of the type (\ref{10.7}) new
possibilities for $\tau$-independent $<\varphi^p(\tau)>$ arise,
corresponding to stationary density matrices in quantum mechanics.

\bigskip
\section{Conclusions and discussion}
\setcounter{equation}{0}
\medskip
Within a simple example of classical statistics for coupled unharmonic
oscillators on a chain we have formulated a description in terms of states and
operators in analogy to quantum mechanics. The state vectors and the operators
can be expressed in terms of classical functional integrals. Typical quantum
mechanical results like the relations between the expectation values in stationary
states or the uncertainty relation can be taken over to the classical system.
Our procedure inverts the construction of the Euclidean path integral for a
quantum mechanical system in the ground state or thermal state, with a generalization
to a wider class of states.

The introduction of ``quantum mechanical'' operators associated to every local classical
observable allows us to define equivalence classes of observables which cannot be
distinguished by any measurement of their expectation values. We argue that the definition
of the correlation function should be consistent with this equivalence structure. We
require that indistinguishable observables must lead to the same correlation function.
This leads to the introduction of a quantum correlation within the classical statistical
setting. We point out that the quantum correlation constitutes a more robust definition
of the correlation function with respect to the precise details of the definition of
observables, both for classical and quantum statistical systems.
The basic conceptual distinction between quantum statistics and classical statistics
disappears in this respect.
We have also seen the emergence of typical characteristics of quantum
statistics like the superposition of states and interference.
All this points into the direction that
it may be possible to understand the mysteries of the basics of
quantum mechanics within a formulation of a classical statistical
problem with infinitely many degrees of freedom.

What is missing is an understanding how the complex structure of quantum mechanics
- the important phases - could be rooted in a possible classical statistical system.
For our present system we discuss in
the two appendices that a complex structure can be defined, if the
system is invariant under a reflection symmetry. There we
introduce a description in terms of normalized complex state vectors
$|\psi(t)>$ which depend on a real-time variable $t$. The time
evolution is then given by the Schr\"odinger equation
$i\partial_t|\psi>=\hat H|\psi>$
with a hermitean
Hamilton operator (\ref{3.11}). Expectation values of observables can be
computed from associated operators in the usual way, e.g.
\begin{equation}\label{11.2}
 <\hat Q^2>=<\psi|\hat Q^2|\psi>\end{equation}
These operators obey
the usual commutation relations, e.g.
\begin{equation}\label{11.3}
[\hat Q,\hat P^2]=2\hat R\quad, \quad [\hat Q,\hat R]=-1\end{equation}
where $\hat R$ plays the role of $i\hat P$.
As familiar in quantum mechanics we can equi\-valently use a
Heisenberg or Schr\"odinger picture for the computation of the time
evolution of expectation values of observables.

We observe, nevertheless, that at the present stage several insufficiencies
remain on our way of understanding quantum mechanics from a classical
statistical formulation. A major problem concerns the restriction that
our description of states as functional integrals is restricted to
those states $|\psi(t)\rangle$ whose analytical continuation
$|\psi(\tau)\}$ for $t=-i\tau$ is a real positive function.
Similarly, we have so far dealt only with linear
superpositions $\alpha_1|\psi^{(1)}\rangle+\alpha_2|\psi^{(2)}\rangle$
with restricted real coefficients $\alpha_i$ such that the sum is again a real
positive function. This situation seems deeply
related to the issue of the complex structure and the
important unsolved questions concerning the
analytic continuation and the role of time.

The quantum mechanical features discussed in this note are present in
all classical statistical systems. However, their conceptual meaning has become
particularly clear in our formulation of incomplete classical statistics. The
approach to incomplete statistics can be extended in various
directions. First of all, incomplete statistical information does not
necessarily occur in the form of missing information outside a local
range. For example, the incompleteness  of the information about the
probability distribution can also concern the resolution within a
given local interval. This problem probably takes a direction which is
qualitatively very different from our development of quantum
mechanics. Within the context used in this paper we may impose
the additional restriction that only information which survives
in the continuum limit $\epsilon\to 0$ is
available. Otherwise stated, the resolution is not arbitrarily
accurate. As discussed in the beginning of sect. 6 the different
derivative observables become indistinguishable in such a version of
incomplete statistics.

One can also investigate the consequences of abandoning
certain of our assumptions. Without the reflection symmetry
$\theta$ we obtain a description
which remains similar to quantum mechanics in many aspects. The
Hamilton operator $\hat H$ does not remain hermitean, however, and the
evolution operator $\hat U$ is not unitary any more (after analytic
continuation). Without translation symmetry in $\tau$ we expect to find
non-zero expectation values of
antihermitean operators. We have not explored what happens if the
interactions go beyond next-neighbor interactions. Finally, our
approach can be extended to a collection of variables $\varphi_a(\tau)$.
If we interpret $a$ as a momentum label, we find a straightforward
generalization to quantum field theory.

At this stage it is still open if the constructions presented here will find some
practical use. Also the crucial question if quantum mechanics can indeed be formulated
within a classical statistical system with infinitely many degrees of freedom has not yet
found a definite answer. We hope that the finding of several structures characteristic
for quantum mechanics within a classical statistical formulation will encourage the
continued exploration in this direction.

\bigskip
\noindent{\bf Acknowledgment:}\\
The author would like to thank M. L\"uscher for useful discussions.

\vspace{1,5cm}
\noindent
\Large \textbf{Appendix}
\normalsize
\appendix
\section{Reflection symmetry and complex structure}
\setcounter{equation}{0}
\renewcommand{\theequation}{A.\arabic{equation}}
\medskip

We have found several quantum-mechanical features in a classical
statistical system, including the characteristic non-commutativity
of operator products. In a sense, the quantum-mechanical
and classical descriptions can be viewed as dual descriptions
of one and the same system, distinguished by the choice of the
correlation function. This raises the question if the other characteristic
features of quantum mechanics, as the complex structure and the unitary
time evolution, can also be found within classical
statistics. In the two appendices we show that these structures
may indeed be present on a formal level. So far we have not found, however,
a fully consistent description of the quantum mechanical phases within our classical
statistical setting.

Quantum mechanics has an intrinsic complex structure. This is manifest
in the Schr\"odinger equation. In this section we establish how this
complex structure emerges from a reflection symmetry in (incomplete)
classical statistics.
Let us consider the reflection\footnote{Existence of the reflection
for every $\tau$ requires that for every variable $\varphi(\tau)$
there is also a variable $\varphi(-\tau)$.}  at a given reference
point $\tau=0$, i.e.
\begin{equation}\label{8.1}
\theta(\varphi(\tau))=\varphi(-\tau),\quad\theta^2=1\end{equation}
We concentrate on reflection-invariant probability distributions \cite{OW}
\begin{eqnarray}\label{8.2}
&&\theta(S[\varphi])=S[\varphi]\nonumber\\
&&\theta({\Lc}(\tau))={\Lc}(-\tau)\nonumber\\
&&\theta(S_>(\tau))=S_<(-\tau)\end{eqnarray}
for which
$V_\tau(\varphi)=V_{-\tau}(\varphi),Z_\tau =Z_{-\tau}$. The
corresponding transformation properties of the states  are
\begin{equation}\label{8.3}
\theta|\psi(\tau)\}=\{\psi(-\tau)|\ ,\quad
\theta\{\psi(\tau)|=|\psi(-\tau)\}\end{equation}
We define  the action of
$\theta$ on matrices such that it also involves a transposition in the
sense that $\theta (\{\psi_1|\hat
A|\psi_2\})=\theta|\psi_2\}(\theta\hat A)\theta \{\psi_1|$. Using the
definitions of $\rho, \hat U,\hat H$ and the operators $\hat A$ one
finds the relations
 \begin{eqnarray}\label{8.4}
&&\theta\rho(\tau)=\rho^T(-\tau)\
, \quad \theta\hat A_{H,S,\delta}(\tau)=\hat
A_{H,S,\delta}^{(R)T}(-\tau),
\nonumber\\
&&\theta\hat H(\tau)=\hat H^T(-\tau)\ , \quad
 \theta \hat U(\tau_2,\tau_1)=\hat U^T(-\tau_1,
-\tau_2)
\end{eqnarray}
Here $\hat A_\delta^{(R)}$ is related to $\hat A_\delta$ by
replacing in eq. (\ref{2.8}) for $A[\varphi(\tau+\eta)
_{-\frac{\delta}{2}<\eta<\frac{\delta}{2}}]$ any $\varphi(\tau+\eta)$
by $\varphi(\tau-\eta)$, corresponding to a reflection of
$A[\varphi,\tau]$ at $\tau$. In consequence of the reflection symmetry
of the probability distribution, the expectation value of any local
observable must be equal to the one of the reflected observable
\begin{equation}\label{8.5}
\theta<A(\tau)>=<A^{(R)}(-\tau)>=<A(\tau)>\end{equation}

On the
other hand, the reflection (\ref{8.1}) acts on $|\psi\}$ only by a
variable charge, i.e.
\begin{equation}\label{8.6}
\theta|\psi(\varphi(\tau);\tau\}=|\psi(\varphi(-\tau));\tau\}=\{\psi(\varphi
(-\tau));-\tau|\end{equation}
Interpreting $|\psi\}$ and $\{\psi|$ as functions
of two variables $\varphi$ and $\tau$ (without distinction to which
$\tau$ the variable $\varphi$ was originally associated) we can write
\begin{equation}\label{8.7}
|\psi(-\tau)\}=\{\psi(\tau)|\end{equation}
and similarly
$\theta(\rho(\tau))=\rho(\tau)$ etc., or
\begin{eqnarray}\label{8.8}
\rho^T(-\tau)=\rho(\tau)&\ ,&\ \hat
A^{(R)T}(-\tau)=\hat A(\tau),\nonumber\\
\hat H^T(-\tau)=\hat H(\tau)
&\ ,&\ \hat U^T(-\tau_1,-\tau_2)=\hat
U(\tau_2,\tau_1)
 \end{eqnarray}
We note that the invariance of the
functional integrals $|\psi\}, \{\psi|,\rho,\hat U,\hat A_\delta$ follows
generally from the possibility to reverse the transformation
(\ref{8.1}) by a variable substitution.  In contrast, the relations
(\ref{8.3}), (\ref{8.4}) involve the invariance properties (\ref{8.2}).

So far, all quantities have been real. We will now introduce a complex
structure by considering two classes of functions of $\tau$ (including operators),
namely those which are either even or odd in $\tau$. Even functions are
considered  as real, whereas the odd ones are purely imaginary. This
relies on the isomorphism between a pair of real functions ($f_{even},
f_{odd}$) and the subclass of the complex functions $z=Re\ z+i\ Im\ z$,
where $Re\ z$ is even and $Im\ z$ is odd, i. e. $\ (Re\ z, i\ Im\
z) \leftrightarrow(f_{even},f_{odd})$. Complex conjugation is then
equivalent to  $(f_{odd}\to -f_{odd})$.
Equivalently, the complex conjugation changes the sign of $\tau$ and
we define its action as
\begin{eqnarray}\label{8.9}
|\psi(\tau)\}^*&=&|\psi(-\tau)\}\ ,\quad
\{\psi(\tau)|^*=\{\psi(-\tau)|,\nonumber\\ \rho(\tau)^*&=&\rho(-\tau)\
,\quad \hat H(\tau)^*=H(-\tau)\nonumber\\ \hat
U(\tau_2,\tau_1)^*&=&\hat U(-\tau_2,-\tau_1)\ , \quad \hat
A(\tau)^*=\hat A(-\tau)\end{eqnarray}
Combining this definition with the action
of the reflection $\theta$ (cf. eqs. (\ref{8.7}), (\ref{8.8})) we
recover well-known properties of quantum mechanics, namely
\begin{eqnarray}\label{8.10}
|\psi(\tau)\}^*&=&\{\psi(\tau)|,\quad\rho^\dagger(\tau)=\rho(\tau),
\quad\hat H^\dagger(\tau)=\hat H(\tau)\nonumber\\
\hat
U^\dagger(\tau_2,\tau_1)&=&\hat U(\tau_1,\tau_2)=\hat U^{-1}(\tau
_2,\tau_1) \end{eqnarray}
With respect to this complex structure the reflection
$\theta$ acts as hermitean conjugation. The euclidean time $\tau$
itself is odd and should therefore be considered as an imaginary
quantity, $\tau=it$, $t$ real.  The self-consistency
of this prescription is apparent by noting that for a wave
function $|\psi(\varphi;\tau)\}=e^{-E\tau}\psi_0(\varphi)$ ($E$ and
$\psi_0$ real) the complex conjugate reads $\{\psi(\varphi;\tau)|
=|\psi(\varphi;\tau)\}^*= (e^{-iEt})^*\psi_0(\varphi)=
e^{iEt}{\psi_0(\varphi)}=
e^{E\tau}\psi_0(\varphi)=|\psi(\varphi;-\tau\}$.  We note that
$|\psi(0)\}=\{\psi(0)|$ and $\rho=\rho^T$ are real.

For an arbitrary operator $\hat A$ we define
\begin{equation}\label{8.11} \hat
A^{(h)}(\tau)=\frac{1}{2}\left(\hat A(\tau) + \hat
A^{(R)}(\tau)\right)\ , \ \hat A^{(a)}(\tau)=\frac{1}{2}\left(\hat
A(\tau)-\hat A^{(R)}(\tau)\right) \end{equation}
From eqs. (\ref{8.8}),
(\ref{8.9}) one finds $\hat A^{(R)}(\tau)=\hat A^\dagger(\tau)$ and we
conclude that $\hat A^{(h)}$ is hermitean and $\hat A^{(a)}$
antihermitean
\begin{equation}\label{8.12} \hat A^{(h)\dagger}(\tau)=\hat
A^{(h)}(\tau),\ \hat A^{(a)\dagger}(\tau)=-\hat A^{(a)}(\tau)
\end{equation}
In
compatibility with eq. (\ref{8.5}) we conclude that all antihermitean
operators must have purely imaginary expectation values. We emphasize
that there is no a priori reason why antihermitean operators should not
be associated with observables. In the sense of the original
definition (\ref{2.4}) their expectation values are simply
odd with respect
to the reflection of $\tau$. The reflection symmetry alone does not
enforce such expectation values to vanish. The situation is different
for simultaneous $\tau$-translation and reflection symmetry where in
addition $<A(-\tau)>=<A(\tau)>$. In this case the expectation values
of all  odd observables or antihermitean operators vanish.

In conclusion, the discrete reflection symmetry introduces the complex
structure characteristic for quantum mechanics: complex state vectors,
hermitean density matrices and a hermitean Hamiltonian. The translation
in $\tau$ is described by a unitary evolution operator.

\bigskip
\section{Analytic continuation}
\setcounter{equation}{0}
\renewcommand{\theequation}{B.\arabic{equation}}

\medskip
The definition of a complex conjugation as an involution in the space
of $\tau$-dependent  functions does not yet specify the complex
structure completely. In addition, we have to define the multiplication with
complex numbers and the complex multiplication of functions. This is
most easily done by constructing a mapping from the space of real
functions of $\tau$ to the space of complex functions where the
operations of complex conjugation and complex multiplication are
implemented in the standard way. We will see that the analytic
continuation of all functions $|\psi(\tau)\}$, $\hat A (\tau)$
etc. constitutes a map with all required properties once $\tau=it$ is
considered as a pure imaginary variable.

We first show that analytic continuation of the functions of interest
is indeed possible in the limit of continuous $\tau$
$(\epsilon\to0)$.  With
\begin{eqnarray}\label{9.1} (\partial/\partial
\tau_2)^p\hat U(\tau_2,\tau_1)&=& (-\hat H(\tau_2))^p\hat
U(\tau_2,\tau_1),\nonumber\\
\ (\partial/\partial\tau_1) ^p\hat
U(\tau_2,\tau_1)&=&\hat U(\tau_2,\tau_1) (\hat H(\tau_1))^p\end{eqnarray}
existing for all $p, \tau_2$ and $\tau_1$ the evolution operator $\hat
U(\tau_2,\tau_1)$ is analytic both in $\tau_2$ and $\tau_1$.  In turn,
$|\psi(\tau)\},\{\psi(\tau)|$ and $\hat A(\tau)$ are analytic
functions of $\tau$. As an example, we may represent $|\psi(\tau)\}$
as a Taylor series with real coefficients $a_n(\varphi)$,
i.e. $|\psi(\varphi;\tau)\}=\sum^\infty_{n=0}
a_n(\varphi)\tau^n$. Similarly, the symmetric and antisymmetric
combinations read
\begin{eqnarray}\label{9.2}
|\psi_s(\tau)\}&=&\frac{1}{2}(|\psi(\tau)\}+|\psi(-\tau)\})\nonumber\\
&=&\sum^\infty_{m=0}a_{2m}(\varphi)\tau^{2m}=
\sum^\infty_{m=0}(-1)^ma_{2m}(\varphi)t^{2m}\nonumber\\
|\psi_a(\tau)\}&=&\frac{1}{2}(|\psi(\tau)\}-|\psi(-\tau)\} \nonumber\\
&=&\sum_{m=0}^\infty a_{2m+1}(\varphi)\tau^{2m+1}=i\sum^{\infty}
_{m=0}(-1)^m a_{2m+1}(\varphi)t^{2m+1}\end{eqnarray}
One sees that the complex
conjugation (\ref{8.9}) is compatible with the analytic continuation
of the real function $|\psi(\tau)\}$ for $\tau\to it$. The
descriptions in terms of the original real functions
$|\psi(\tau)\}=|\psi_s (\tau)\}+|\psi_a(\tau)\}$ or the complex wave
functions $|\psi(it)\}=|\psi_s(it)\}+|\psi_a(it)\}$ are completely
equivalent -- they are simply related to a change of variables. We
repeat, however, that the use of the complex structure related to
$\tau$-reflection necessarily implies that $\tau$ must be purely
imaginary. (Real values for $\tau$ are not compatible with this
complex structure and should be used only in the language employed
originally where all quantities are real. Obviously, the meaning of
the word ``real'' depends on the complex structure used to distinguish
between real and imaginary numbers. In absence of a complex structure
all quantities are trivially ``real''.)

The complex multiplication of two functions $|\psi_1(it)\}$ and
$|\psi_2(it)\}$ is equivalent to the (real) multiplication of
$|\psi_1(\tau)\}$ and $|\psi_2(\tau)\}$ in the original language
\begin{equation}\label{9.3}
|\psi_3(\tau)\}=|\psi_1(\tau)\}|\psi_2(\tau)\}
\leftrightarrow |\psi_3(it)\}=|\psi_1(it)\}|\psi_2(it)\}\end{equation}
This is a
direct consequence of the compatibility of analytic continuation with
the product of two complex functions.  The multiplication law
(\ref{9.3}) extends to  matrix products $\hat A_1(it)\hat A_2(it)$ or
$\hat U(it_2,it_1)|\psi(it_1)\}$.  In the following we will always use
the complex structure with the complex multiplication and adopt the
notation
\begin{eqnarray}\label{9.4}
&&|\psi(t)>\equiv|\psi(it)\},\ <\psi(t)|\equiv
\{\psi(it)|=|\psi(-it)\}= |\psi(t)>^*\nonumber\\
&&U(t_2,t_1)\equiv\hat U(it_2,it_1),\ A^{(M)}(t)\equiv\hat A(it)\end{eqnarray}
In other words, $|\psi(t)>$ is the analytic continuation of
the functional integral
$|\psi(\tau)\}$ given by eq. (\ref{2.7}). In particular, for an exponentially decreasing
$|\psi(\tau)\}=e^{-E\tau}\psi_0$ the state vector
$|\psi(t)>=e^{-iEt}\psi_0$ is an oscillatory complex function of
$t$. We  note that $U$ is unitary
\begin{equation}\label{9.5}
U^\dagger(t_2,t_1)U(t_2,t_1)=1\end{equation}
and the time evolution conserves the
norm of the complex state vector
\begin{equation}\label{9.6}
<\psi(t)||\psi(t)>=\{\psi(\tau)||\psi(\tau)\}=1\end{equation}
We associate the variable $t$ with time and recover the Schr\"odinger
equation (cf. (\ref{3.2}))
\begin{equation}\label{9.7}
i\partial_t|\psi>=H|\psi>\end{equation}

At this point we should remark that our one-dimensional
classical statistical system with next-neighbor interactions
(\ref{1.1}) has not a priori an interpretation as a description
of a time evolution. In fact, one possibility is to view
this system as a chain in space in equilibrium. In this case
it describes infinitely many interacting degrees of freedom
and there is no time evolution. The quantum mechanical discreteness
is present, nevertheless, as well as the description in terms of operators
and the possibility to introduce a complex structure. We have seen
in appendix A that we can define operators and correlation
functions in dependence on a complex parameter $\eta=e^{i\alpha}t$,
e.g.
\begin{equation}\label{9.8}
\hat Q_H(\eta)=e^{i\eta\hat H}\ \hat Q\ e^{-i\eta\hat H}\end{equation}
Using the complex language (which is not familiar in this context)
the relevant correlation functions for a chain in
space correspond to purely imaginary $\eta$. Actually, the concepts of
distance and geometry (in this case trivial) can be constructed from
the $\eta$-dependence of the correlation function without assuming
a space interpretation a priori \cite{Geom}.

A different interpretation of the probability distribution
(\ref{1.1}) concentrates on expectation values and correlations
for real $\eta$. In this case $\eta$ can be considered as a time
parameter and our system describes the quantum-mechanical
time evolution of a single degree of freedom (not infinitely
many as for the space interpretation). We emphasize that
this interpretation
is not characterized by a given deterministic  time
evolution equation. The concept of time arises from the
``transport of information'' between neighboring regions (``time
regions'' in this case) and is itself of a probabilistic
nature! In this sense we may consider the dynamics of quantum
mechanics as the result of an interpretation of time as a particular
structure among observables in a classical statistical system.

Our definition of a complex structure within incomplete
statistics allows the computation of expectation values of
time-dependent observables $A(t)$ for a range of complex values of
$t$. (The imaginary part of $t$ should be within the local region
$|\tau|\leq\bar\tau$
introduced in the formulation of our problem.)
Although suggestive, it is not
apparent in the present framework why our interpretation of
the real world concentrates on real values of $t$.
At the present stage it is not obvious
why the values of the correlation functions
at imaginary values of $\tau$ (real $t$) are much
more important than their values of real $\tau$.
We believe that a satisfactory answer to these questions
will shed more light on the basic origins of quantum mechanics and
time.

If $\tau$ becomes a complex variable we may also consider analytic continuation
directly in the functional integral. This can easily be achieved by choosing a purely
imaginary $\epsilon$ in eqs. (\ref{2.2}) and (\ref{2.3}). Such a prescription brings us,
however, to a path integral with phases for which the positivity of $\exp (-S)$ and
therefore the classical probabilistic interpretation is lost. On the other hand, we
could now release the restriction that the analytic continuation of $|\psi(t)>$ leads
to a positive real function $|\psi(\tau)\}$, opening the possibility to describe excited
states. At the present stage we have not yet succeeded to overcome this conflict
between the phases of quantum mechanics and the probabilistic interpretation of
$\exp -S$. It seems worthwhile to investigate if there could exist more complex
statistical systems for which the analytical continuation from euclidean to real time
is consistent with a probabilistic interpretation.

\end{document}